\documentclass[%
 reprint,
 amsmath,amssymb,
 aps,
]{revtex4-2}

\usepackage{graphicx}
\usepackage{dcolumn}
\usepackage{bm}
\usepackage{color}
\usepackage{float}

\begin{document}

\preprint{APS/123-QED} 

\title{A gravitational decoupling MGD model in modified $f(R,T)$ gravity theory}

\author{S. K. Maurya} \email{sunil@unizwa.edu.om}
 \affiliation{ Department of Mathematics and Physical Science,
College of Arts and Science, University of Nizwa, Nizwa, Sultanate of Oman}
\author{Abdelghani Errehymy}\email{abdelghani.errehymy@gmail.com} \affiliation{Laboratory of High Energy Physics and Condensed Matter (LPHEMaC), Department of Physics, Faculty of Sciences A\"{i}n Chock, University of Hassan II, B.P. 5366 M\^{a}arif, Casablanca 20100, Morocco}
\author{Ksh. Newton Singh} \email{ntnphy@gmail.com}
\affiliation{Department of Physics, National Defence Academy, Khadakwasla, Pune 411023, India \\
Department of Mathematics, Jadavpur University, Kolkata-700032, India}
\author{Francisco Tello-Ortiz} \email{francisco.tello@ua.cl}
\affiliation{Departamento de F\'isica,
Facultad de ciencias b\'asicas, Universidad de Antofagasta, Casilla 170,
Antofagasta, Chile}
\author{Mohammed Daoud} \email{m$_{}$daoud@hotmail.com} \affiliation{Department of Physics, Faculty of Sciences, University of Ibn Tofail, B.P. 133, Kenitra 14000, Morocco\\
Abdus Salam International Centre for Theoretical Physics, Miramare, Trieste 34151, Italy}

\date{\today}

\begin{abstract}
The present paper is devoted to investigating the possibility of getting stellar interiors for ultra-dense compact spherical systems portraying an anisotropic matter distribution employing the gravitational decoupling by means of Minimal Geometric Deformation (MGD) procedure within the modified theory of $f(R,T)$ gravity. According to this theory, the covariant divergence of stress-energy tensor does not vanish, hence the movement of classical particles does not follow geodesics resulting in an extra acceleration which suffices the late-time acceleration of the universe without adopting to exotic matter fields. In this regard, we have considered the algebraic function as $f(R, {\rm T})= R+2\chi T$, the corresponding effective stress-energy tensor is conserved as well as the exact solutions are derived, where $\chi$ indicates a coupling constant. Moreover, the physical quantities associated with the new solutions are well-behaved from the physical and mathematical point of view as well as free of geometrical singularities, violation of the causality condition, non-decreasing thermodynamic functions. Thereafter, the physical viability of the obtained model is affirmed by performing several physical tests of the main salient features such as energy density, radial, and tangential pressure, anisotropy effect, dynamical equilibrium, energy conditions, and dynamical stability. On the other hand, we have generated the $M-R$ curves from our solutions in the four different scenarios, including GR, GR+MGD,  $f(R,T)$ and $f(R,T)+$MGD, and we found a perfect fit for many compact spherical objects in these scenarios by changing the gravitational decoupling constant $\alpha$ and the coupling constant $\chi$ as free parameters. The present study reveals that the modified $f(R,T)$ gravity through gravitational decoupling by means of MGD method is a suitable theory to explain compact stellar spherical systems like, X-ray binaries viz., Vela X-1, Cen X-3, Cyg X-2, LMC X-4, 4U 1538 52 and Her X-1, low mass X-ray binaries viz., 4U 1820-30 and 4U 1608-52, and binary millisecond pulsars viz., PSR J16142230 and PSR J1903+327, and many more another compact stellar spherical systems respecting the well-known and tested general requirements.
\end{abstract}

\maketitle

\section{Introduction}\label{sec1}
Einstein's famous theory of gravity comprises of reasonable creativity and scientific style in its each progression. To represent non-inertial systems and floating structures from a standard standpoint at large scale phases, Einstein's gravity theory is a very helpful implement \cite{Einstein:1915a}. However, despite its excellence, the geometrical singularities make it stationary in certain situations \cite{Wheeler:1962a}. Moreover, the theory cannot continue normally for the representation of the period of observational acceleration of the Universe. Cannot adequately depict the enormous elements of the Universe without taking into consideration the exotic type of energy-matter that should be known as the dark side (dark-energy and dark-matter) \cite{Spergel:2007a,Caldwell:2002a,Nojiri:2003a,Riess:2004nr, Eisenstein:2005a,Nojiri:2003ab, Errehymy:2017, Errehymy:2019, Astier:2006a, Kamenshchik:2001a, Padmanabhan:2002a, Bento:2002a}. Einstein's theory of gravity does not account into the matter quantum kind and cannot be quantified in a renormalization traditional method. In this reasoning, it has been indicated \cite{Utiyama:1962a} that taking into account of higher-order curvature and flow terms in Einstein-Hilbert's action will make it renormalizable framework into a loop. Then again, to incorporate quantum adjustments, the higher-order curvature and flow invariants in the effective gravitational action at the low energy phase must be considered \cite{Vilkovisky:1992a, Birrell:1982ab, Buchbinder:1992a}. In spite of the fact that the theory stood the trial of time and has various applications in physical cosmology, it can't clarify the most serious issue in physical cosmology, i.e., the kind of cosmological constant. The acceleration of the universe at the present age can't be clarified without summoning new types of matter as well as energy \cite{Pavlovic:2017ab}. There are two possible cases to address this exceptional issue: the first case, numerous candidates for dark energy have been suggested in the literature, such as f-essence, k-essence, quintessence, spintessence, tachyons field, ghost field, Chaplygin gas \cite{Copeland:2006ab}. In the second case, we can persuade ourselves that we are living in a Universe with a cosmological constant or scalar field that accelerates the Universe on a larger scale. There are numerous studies on the nature of dark energy and cosmic expansion based on different methods. These researches can be organized as follows: to modify all cosmic energy by including new elements of dark energy and to modify the action of Einstein-Hilbert to obtain various types of modified theories that include $f(R)$ gravity \cite{Capozziello:2008a,Capozziello:2009a,Nojiri:2009a,Felice:2010a,Maartens:2010a,Capozziello:2010a,Capozziello:2011a,Nojiri:2011a,Capozziello:2012a,Astashenok:2013a,Astashenok:2015a,Capozziello:2015a,Astashenok:2015b,Jovanovic:2016a,Capozziello:2016b,Santos:2017a,Astashenok:2017a,Chervon:2018a,Capozziello:2018a,Capozziello:2018b,Capozziello:2018c,Odintsov:2019a,Odintsov:2019b,Capozziello:2019a}, $f(\mathbb{T})$ gravity \cite{Bohmer:2011a,Wang:2011a,Daouda:2011a,Sharif:2013a,Capozziello:2013b}, $f(R,T)$ gravity \cite{Harko:2011a,pop,Yousaf:2016a,Yousaf:2016b,Correa:2016a,Moraes:2017a,Das:2016a,Deb:2018a,Maurya:2019ab,Deb:2019a,Sahoo:2019a,Shabani:2018a,Wu:2018a,Barrientos:2018a,Hansraj:2018a,Singh:2018a,Baffou:2018a,Hansraj:2018b,Yousaf:2018b}, $f(\mathbb{T})$ gravity \cite{Bengochea:2009a,Linder:2010b}, $f(\mathbb{T},T)$ gravity \cite{Momeni:2014a,Junior:2015a,Nassur:2015a,Salako:2015a,Saez-Gomez:2016a,Pace:2017a,Lobo1,Lobo2}, $f(\mathcal{G})$ gravity \cite{Bamba:2010a,Bamba:2010b,Rodrigues:2014a}, $f(R, \mathcal{G})$ gravity \cite{Nojiri:2005a}, where $R$, $T$, $\mathbb{T}$ and $\mathcal{G}$ are Ricci's scalar, trace of stress-energy tensor, torsion scalar and Gauss-Bonnet scalar respectively. The $f(R)$ theory is an appropriate theory that modifies Einstein's gravity with the substitution of action $R$, by a general expression of scalar curvature $R$  \cite{Capozziello:2002b,Carroll:2004a,Dolgov:2003a,Nojiri:2003b,Nojiri:2004c,Copeland:2006a}. Starobinsky \cite{Starobinsky:1980a} assumed a cosmological model of inflation employing $f(R)$ gravity. Starobinsky \cite{Starobinsky:2007a}, Hu and Sawicki \cite{Hu:2007a} inserted cosmological models employing nonlinear expressions $f(R)$.

In the same spirit, Capozziello and his accomplices \cite{Capozziello:2014c} exhibited a review of $f(R)$ gravity and talked about the clarification of dark energy and inflation periods, depicting the quickened periods of the Universe, with regards to $f(R)$ gravity. They likewise depicted some significant constituents of inflation, cosmography and quintessence in the context of $f(R)$ gravity theory. Cruz-Dombriz and his partners \cite{Cruz-Dombriz:2016a} figured the limits on model variables of modified theories that include quintessence, $f(R)$ and $f(\mathbb{T})$ and utilized cosmography as an instrument to separate these theories. Furthermore, Aviles et al. \cite{Aviles:2013a} examined the development of the appropriate type of gravity theory $f(\mathbb{T})$ in a cosmographic manner. The first associated the capacity $f(\mathbb{T})$ and it's differential coefficients with the cosmographic variables and fixed the cosmographic restrictions which are free of the model and estimated that the obtained outcomes were reliable with another cosmological model. D'Agostino and Luongo \cite{D'Agostino:2013a} considered teleparallel dark energy simultaneously with a non-minimal coupling among the scalar field and torsion and contrasted this situation with that of the quintessence. They exhibited that this energy has an enormous set of arrangements than controlled by minimal quintessence. For this purpose, the same authors investigated the rate of perturbation development and contrasted the outcomes with another cosmological model.

In addition, the treatment of related open research in the cosmological domain is also a promising methodology \cite{Felice:2010a} that is closely related to the $f (R, T)$ theory. This theory is created by considering non-minimal coupling among the trace of the stress-energy tensor $T$ and Ricci Scalar $R$. Harko and his colleagues \cite{Harko:2011a} acquainted it first to handle the problems in a proficient manner, introducing a new modification of Einstein's theory of general relativity and the $f(R,T)$ theory of gravity. In this $f(R,T)$ theory of gravity, the matter is well-respected on an equivalent equilibrium with configuration. Thus one can investigate numerous fascinating and novel highlights of the Universe, for example, the dark matter part \cite{Gibbons:1977a}. One can take note of that reliance of $T$ may originate from the thought of quantum impacts or from the nearness of a flawed fluid. The concept of coupling of the matter and geometry fields in a non-minimal manner returns to Rastall \cite{Rastall:1972a}, who tested the stress-energy conservation law in the bended spacetime just because, a hypothesis upheld by the particle creation during the Universe development \cite{Gibbons:1977a,Parker:1971a,Ford:1987a,Birrell:1982a,Brax:2007a}. Along these lines, it was appeared by Smalley \cite{Smalley:1984a} that a model of Rastall's gravity, where the uniqueness of the stress-energy tensor is relative to the Ricci scalar gradient, can be logical from a variational rule.

Besides, some static spherically symmetric arrangements of the Rastall field equations have been acquired. For instance, by uncommon adjustment of the parameters of the Rastall hypothesis, a vacuum arrangement having a similar configuration as the Schwarzschild-de Sitter arrangement in the theory of general relativity acquired with cosmological constant acting as a source \cite{Oliveira:2016a}. In this regard, Bronnikov et al \cite{Bronnikov:2016a} examined the problem of finding static spaces with spherical symmetry in Rastall theory in the presence of a free scalar field or with self-interaction. It was discovered that it is possible to obtain specific arrangements in which part of these arrangements are equivalent to the arrangements that are acquired with respect to the k-essence hypothesis. Furthermore, asymptotically level through-gap arrangements are examined in \cite{Moradpour:2016a} where it is demonstrated that the Rastall hypothesis parameters influence the wormholes space-time parameters.

Nevertheless, despite its great versatility recently as claimed by H. Velten and T. R. P. Caram$\hat{\text{e}}$s \cite{Hermano:2017} the Cosmological inviability of $f(R,T)$ gravity theory.
In their work they tested the general functional
\begin{equation}\label{functional}
f(R,T)=R+\lambda T+\gamma_{n}T\,^{n},    
\end{equation}
where $R$ is the Ricci scalar, $T\equiv g^{\mu\nu}T_{\mu\nu}$ the trace of the energy-momentum tensor, $\lambda$, $\gamma$ and $n$ constants with $n$ restricted to be $n\geq 0$. It was shown by the authors that the general model (\ref{functional}) for different values of the parameter $n$ does not match the observational data. Hence, it is not possible to explain the existences of dark components \i.e, dark energy and dark matter. Of course, this antecedent reduce the utility of $f(R,T)$ gravity theory as an alternative to GR to explain the accelerated era of the Universe. However, the general functional given by (\ref{functional}) and others models with logarithmic form containing the contribution of the trace of the energy-momentum tensor have been used to study the feasibility in the line of compact structures describing stellar interiors in the arena of $f(R,T)$ theory. In this regards, the formulation of a suitable solution of a spherically symmetrical self-gravitating structure has still been difficult due to the presence of non-linearity in the field equations. Many studies have been made in the literature to address this problem. Mak and Harko \cite{Mak:2004a} acquired a precise anisotropic disposition of the field equations and establish a positively finished performance of pressure and density that contribute to the nucleus of astrophysical objects. Gleiser and Dev \cite{Gleiser:2004a} studied anisotropic self-gravitating frame algorithm with a mass-to-radius ratio $M/R=4/9$ and acquired steady model for small estimates of the adiabatic index. Kalam and his colleagues \cite{Kalam:2013a} discovered compact models with regards to the anisotropic system utilizing Krori and Barua metric. Maurya et al. \cite{Maurya:2017a} examined compact stars anisotropic solutions in terms of charge distribution. Moreover, in \cite{Deb:2019,Maurya:2019b} were obtained compact objects representing real celestial bodies such as neutron and quark stars. These solutions were study within the background of $f(R,T)$ gravity theory by employing the so called embedding class I approach or Karmarkar condition \cite{Karmarkar:1948}. 

The existence of accurate inside solutions of self-gravitating frameworks within the sight of anisotropy has been achieved in different manners. In this respect, the Minimal Gravitational Decoupling (MGD) procedure showed up as importantly settled in finding the physically feasible answers for spherically symmetric astrophysical geometry. Ovalle \cite{Ovalle:2008b} used this methodology to find precise answers to compact astrophysical objects as part of Braneworld. However, the complete details of the MGD approach in the Brane-World problem is given in Ref.  \cite{Ovallebook}. Moreover, this MGD approach is a very powerful technique for exploring new spherically symmetrical solutions of Einstein's field equations. The principal property of this method is that a straightforward solution can be prolonged to more complicated areas. Ovalle and his collaborators \cite{Ovalleb,Ovallec,Ovalled,Ovallee,Ovallef,Ovalleg,Ovallei,Ovallej,Ovallek} have done pioneering works in the context of MGD as well as in the extended MGD scenario.  The starting point of this procedure is to couple to the seed energy-momentum tensor $T^{(\text{m})}_{\mu\nu}$ representing a perfect fluid (this represents the simplest case, however it can be anisotropic also) an additional source $\theta_{\mu\nu}$. This is done through a dimensionless parameter, namely $\alpha$. So, the extended or total energy-momentum tensor now reads $T^{(m)}_{\mu\nu}$~$\longrightarrow$~$T^{(tot)}_{\mu\nu}$ = $T^{(m)}_{\mu\nu}$ + $\alpha \theta_{\mu\nu}$. The $\alpha$ parameter plays an important role into the feasibility of describing anisotropic compact structures. It sign and magnitude depends on the mechanism selected to close the $\theta$-system in order to determine the $\theta_{\mu\nu}$ components and the decoupler function. In this respect, one has several ways to face the closure of the set of equations corresponding to the additional source $\theta_{\mu\nu}$. For example in the seminal work \cite{Ovalle:2018c} the mimic constraint procedure was proposed and employed to deform Tolman IV solution, what is more the same technique was used in \cite{Gabbanelli:2018}, \cite{Tello:2018} and \cite{Morales:2018aa} to extend the Durgapal-Fuloria, Heintzmann IIa and Charged Heintzmann IIa solution respectively, to an anisotropic domain. Basically, this approach consists of matching the isotropic  pressure $p$ with the $\theta^{r}_{r}$ component or the seed energy-density $\rho$ with the temporal component $\theta^{t}_{t}$, arriving to an algebraic equation in the former or a differential equation in the second case. It should be noted that all this explanation is within the framework of GR. In the case of modified gravity theories the resulting equations leading to the determination of the deformation function $f(r)$ can be different. So, this way yields to find the concrete form of the minimal deformation function $f(r)$. Nevertheless, as was shown in \cite{Morales:2018a,Maurya:2019a} also it is possible to impose an adequate form for $f(r)$ respecting all the physical and mathematical requirements in order to close the $\theta$-system of equations. So, in the previous works by using the mimic constraint procedure naturally a lower and upper bounds appear when $p=\theta^{r}_{r}$, namely $0<\alpha<1$. This restriction on $\alpha$ are to ensure a positive defined anisotropy factor $\Delta$ everywhere. In the case when $\rho=\theta^{t}_{t}$ to guarantees $\Delta>0$ the parameter $\alpha$ is allowed to take negative values. On the other hand, the methodology following in \cite{Morales:2018a,Maurya:2019a} is not too much restrictive in the choice of $\alpha$ (sign and magnitude), because it depends on the behaviour of the model.

In these circumstances, Casadio and his partners \cite{Casadio:2015a} implemented external responses to the self-gravitating frame with spherical symmetry using the gravitational decoupling method and established a geometry of naked singularity at the Schwarzschild radius. Ovalle et al \cite{Ovalle:2018c} expanded the isotropic arrangement by incorporating an anisotropy employing the MGD technique for a static astrophysical object system. According to this technique, Sharif and Sadiq \cite{Sharif:2018c} studied a charged anisotropic spherical solution and further analyzed the stability criteria based on sound velocity, availability conditions.

In the present article the gravitational decoupling by means of MGD is combined with the $f(R,T)$ gravity methodology to investigate the the possibility of obtaining in analitycal way compact structures describing real static and spherically symmetric self-gravitating configurations such as neutron or quark stars. As we will see, the general field equations in the picture of $f(R,T)$ gravity theory are more complicated than the GR ones. So, the gravitational decoupling by MGD approach depends on the election of the $f(R,T)$ functional. In this respect we have selected $f(R,T)=R+2\chi T$, being $R$ the Ricci scalar, $T$ the trace of the energy-momentum tensor and $\chi$ a dimensionless constant. This simple choice allows to decouple the $f(R,T)$ sector from the $\theta$-sector, what is more the $f(R,T)$ contribution is carried out into the $\theta$-sector contributing in the final form of the decoupler function $f(r)$. As pointed out above, in \cite{Hermano:2017} was proved that the model given by (\ref{functional}) is not viable from the cosmological point of view (this model matches our election when $\gamma_{n}=0$). Nevertheless, the failure to explain the problems at the cosmological level does not imply that $f(R,T)$ gravity theory should be discarded from the astrophysical plane. In the literature there are some novel works devoted to the study of compact structures within the arena of $f(R,T)$ gravity theory employing a linear functional in $R$ and $T$, showing the feasibility of the model. For example in \cite{fr1} $f(R,T)$ filed equations were solved numerically finding that the  maximum mass allowed for neutron and strange stars are  $1.538 M_{\odot}$ and $2.017 M_{\odot}$ respectively. 
Furthermore, in\cite{fr2,fr3,fr4,fr5,fr6}
were considered the inclusion of different ingredients such as electric charge and anisotropies to solve the problem in analitycal form.  Furthermore, the study of collapsed configurations using different models for the $f(R,T)$ functional and investigations on exotic objects such as gravastar can be found at \cite{fr7,fr8}. Hence, all these good antecedents motivate and support the present investigation. Moreover, gravitational decoupling by MGD has shown to be a powerful and versatile method to work out in different context. Therefore, despite the simple choice considered by us about the $f(R,T)$ functional, this work presents a first step in the understanding of how to decouple gravitational sources which represent analytical models for the description of real celestial bodies. What is more, in the limit $\{\alpha\rightarrow0,\chi\rightarrow0\}$ general relativity results are recovered, hence the obtained macro-physical observables such as the the mass $M$ and radius $R$ as well as the central data of the thermodynamic variables driven the micro-physical process, namely $\{\rho(0),p_{r}(0)\}$ can be compared with those provided by Einstein theory to check the possibility of exceeding the limits established for these quantities within the framework of classical theory of gravity.

So, we have organized the present paper as follows: fundamental mathematical detailing of $f(R,T)$ gravity is introduced in Section \ref{sec2}. In Section \ref{sec3} we formulate fundamental stellar equations in $f(R,T)$ gravity theory and present the solution of the Einstein field equations for multiple sources. In Section \ref{sec4} we discuss the gravitational decoupling by the MGD approach in the $f(R,T)$ gravity structure, and present the matching conditions of our stellar model in Section \ref{sec5}. In Section \ref{sec6} we incorporate the decoupler MGD procedure, so as to get some of the limitations imposed on the field equations, namely specific the mimic constraints. Section \ref{sec7} discusses the ansatz for the metric function $g_{rr}$ suggested by Korikina and Orlyanskii according to our solutions. Section \ref{sec8} We examine the dynamical equilibrium of the stellar system. Then in Sections \ref{sec9}, \ref{sec10} and \ref{sec11} we analysis the physical acceptability and stability of the stellar system, by investigating energy conditions, causality condition, surface redshift, adiabatic index and fitting of observed values in $M-R$ curves is given in Section \ref{sec12}. Finally, the concluding remarks close this paper in the last section.

\section{$f(R, T)$ gravity formalism}\label{sec2}

Let us present $f(R, T)$ gravity. We introduce the modified gravity theory and its extension to $f(R, T)$ theory. The complete action in this theory is given by
\begin{eqnarray}\label{eq1}
S &=& \frac{1}{16\pi}\int f(R,T)\sqrt{-g}~ d^{4}x+\int \mathcal{L}_{m}\sqrt{-g}~ d^{4}x\nonumber\\
&& +\alpha\int \mathcal{L}_{\theta}\sqrt{-g}~ d^{4}x. 
\end{eqnarray}
Here the generic function $f(R,T)$ contains the trace of the stress-energy tensor $T$ as well as $R$, the Ricci scalar, $\mathcal{L}_m$ is the matter Lagrangian density which represents the possibility of non-minimal coupling between geometry and matter, $\mathcal{L}_{\theta}$ is the Lagrangian density of a new sector and $g$ stands for the determinant of the metric $g_{\mu\nu}$. In this respect, the stress-energy tensor is described by the following expressions

\begin{eqnarray}
T_{\mu\nu}=-\frac{2}{\sqrt{-g}}\,\frac{\delta(\sqrt{-g}\,\mathcal{L}_m)}{\delta g^{\mu\nu}}, \label{eq2}\\
\theta_{\mu\nu}=-\frac{2}{\sqrt{-g}}\,\frac{\delta(\sqrt{-g}\,\mathcal{L}_\theta)}{\delta g^{\mu\nu}}.\label{eq3}
\end{eqnarray}
Following Harko et al. \cite{Harko:2011a} and using $T=g^{\mu\nu}T_{\mu\nu}$ with respect to the situation of the matter Lagrangian density $\mathcal{L}_m$ depends solely on the metric tensor $g_{\mu\nu}$, the stress-energy tensors expressed in Eqs. (\ref{eq2}) and (\ref{eq3}) can be written as follows
\begin{eqnarray}
T_{\mu\nu} &=& g_{\mu\nu} \mathcal{L}_m-\frac{2\,\partial(\mathcal{L}_m)}{\partial g^{\mu\nu}}, \label{eq4}\\
\theta_{\mu\nu} &=& g_{\mu\nu} \mathcal{L}_\theta-\frac{2\,\partial(\mathcal{L}_\theta)}{\partial g^{\mu\nu}}. \label{eq5}
\end{eqnarray}

The full action (\ref{eq1}) can be varied regarding the metric tensor $g^{\mu\nu}$ to acquire the general gravitational field equations for $f(R, T)$ gravity as 
\begin{eqnarray}\label{6}
\left( R_{\mu\nu}- \nabla_{\mu} \nabla_{\nu} \right)f_R (R,T) +\Box f_R (R,T)g_{\mu\nu} - \frac{1}{2} f(R,T)g_{\mu\nu}   \nonumber \\ = 8\pi\,(T_{\mu\nu}+\alpha\,\theta_{\mu\nu}) - f_T (R,T)\, \left(T_{\mu\nu}  +\Theta_{\mu\nu}\right),~~~~~
\end{eqnarray}

where $f_R (R,T)={\partial f(R,T)}/{\partial R}$, $f_T (R,T)={\partial f(R,T)}/{\partial T}$ and $\nabla_\mu$ is the covariant derivative related to the Levi-Civita connection of metric tensor $g_{\mu\nu}$ and  the D'Alambert operator $\Box$ which is defined as

\begin{eqnarray}
\Box\equiv\partial_\mu(\sqrt{-g}~g^{\mu\nu}\partial_\nu)/\sqrt{-g}, ~~ \textrm{and} ~~ \Theta_{\mu\nu}=g^{\mu\nu}\delta T_{\mu\nu}/\delta g^{\mu\nu}. \nonumber
\end{eqnarray}
In obtaining the covariant derivative expression of the stress-energy tensor as well as the algebraic function extract, the covariant derivative of formula (\ref{6}) is performed as follows
\begin{eqnarray}\label{eq7}
\nabla^{\mu}T_{\mu\nu}&=&\frac{f_T(R, T)}{8\pi -f_T(R,T)}\bigg[(T_{\mu\nu}+\Theta_{\mu\nu})\nabla^{\mu}\ln f_T(R,T)  \nonumber \\&+&\nabla^{\mu}\Theta_{\mu\nu}-\frac{1}{2}g_{\mu\nu}\nabla^{\mu}T-\frac{8\pi\,\alpha}{f_{T}\left(R,T\right)} \nabla^{\mu}\theta_{\mu\nu}\bigg].
\end{eqnarray}

It is obvious to see that the stress-energy tensor $T_{\mu\nu}$ in $f(R,T)$ gravity theory is not preserved, as in other modified gravity theories \cite{Yu:2018a,Zhao:2012a}. Now using Eq. (\ref{eq4}), we define the tensor formula $\Theta_{\mu\nu}$ as follows
\begin{equation}\label{8}
\Theta_{\mu\nu}= - 2 T_{\mu\nu} +g_{\mu\nu}\mathcal{L}_m - 2g^{\alpha\beta}\,\frac{\partial^2 \mathcal{L}_m}{\partial g^{\mu\nu}\,\partial g^{\alpha\beta}}.~~~
\end{equation}

To describe inside spacetime of the spherically symmetric and static astrophysical structure in Schwarzschild-like coordinates ($t, r, \theta, \phi$), we use line element as follows
\begin{equation}\label{9}
ds^{2} = -e^{\nu(r) } \, dt^{2}+e^{\lambda(r)} dr^{2}+r^{2}(d\theta ^{2} +\sin ^{2} \theta \, d\phi ^{2}),
\end{equation}
where $e^{\nu(r)}$ and $e^{\lambda(r)}$ depict the gravitational potentials of astrophysical configuration. In the further investigation, we adopt the geometrical units $G=c=1$. We suppose that the inside of spherical object is filled of an ideal fluid source, in the current study we are taking into consideration the stress-energy tensor in the following form
\begin{equation}\label{10}
T_{\mu\nu}=(\rho+p) u_\mu u_\nu-p g_{\mu\nu},
\end{equation}
where the covariant component ${u _{\nu}}$ denote the 4-velocity, fulfilling $u_{\mu}u^{\mu}= -1$ and $u_{\nu}\nabla^{\mu}u_{\mu}=0$.
Here, $\rho$ and $p$ represent the pressure and matter density for isotropic matter. In the current investigation, according to the definition proposed by Harko et al. \cite{Harko:2011a}, we assume $\mathcal{L}_m= -p$ and applying the tensor (\ref{8}) we find
\begin{eqnarray} \label{eq11}
\Theta_{\mu\nu}=-2T_{\mu\nu}-p\,g_{\mu\nu}.
\end{eqnarray}

The algebraic function $f(R, T)$ can be chosen in many ways corresponding to feasible models. In the present work, we have considered the algebraic function as 
\begin{equation}\label{eqq12}
f(R, T)= R+2\chi T
\end{equation}
in order to establish the effective energy-momentum tensor ${T}^{eff}_{\mu\nu}$, where $\chi$ indicates a coupling constant. This choice of $f(R,T)$ can be use to resolve the cosmological constant problem, where the cosmological constant is represented by an effective value proportional to Hubble function ($H=\dot{a}/a$) i.e. $\Lambda_{eff} \propto H^2$ \cite{Harko:2011a,pop}. By using this linear function and with the help of the general function of motion (\ref{6}), Einstein's tensor is explicitly reduced as
\begin{eqnarray}\label{eq12}
G_{\mu\nu}= 8\pi\,(T_{\mu\nu}+\alpha\,\theta_{\mu \nu})+\chi T g_{\mu\nu}+2\chi(T_{\mu\nu}+p\,g_{\mu\nu})\nonumber\\=8\pi \tilde{T}_{\mu\nu}+8\pi\,\alpha\,\theta_{\mu \nu}={T}^{eff}_{\mu\nu}.~~~
\end{eqnarray}

It should be noted that when the algebraic function reduces to curvature scalar $R$, the field equations (\ref{6}) turning directly to Einstein field equations. Examining such specific linear supposition has generally acknowledged addressing cosmological as well as stellar solutions. By exchanging the value of the functional $f(R, T)$ in Eq. (\ref{eq7}), we get

\begin{eqnarray}\label{eq13}
\nabla^{\mu}T_{\mu\nu}=-\frac{1}{2\,\left(4\pi+\chi\right)}\chi\bigg[g_{\mu\nu}\nabla^{\mu}T+2\,\nabla^{\mu}(p g_{\mu\nu})\nonumber\\+\alpha~ \frac{8\pi}{\chi}\nabla^{\mu}\theta_{\mu\nu}\bigg].
\end{eqnarray}

However, this Eq. (\ref{eq13}) can be written in the following achievable form $\nabla^{\mu}{T}^{eff}_{\mu\nu}$=$0$, which depicts the conservation of the stress-energy tensor ${T}^{eff}_{\mu\nu}$ for the effective distribution of the matter. Note that if we disable the coupling $\alpha$ and the free parameter $\chi$ takes the value zero, the standard results due to the general relativity can be achieved.


\section{Basic stellar equations in $f(R, T)$ gravity theory}\label{sec3}
As we are focused on examining a spherical fluid, we consider the static, spherically symmetric astrophysical structure, for which the most general line element of the inside spacetime, can be depicted as the following metric
\begin{equation}\label{eq14}
{ds}^{2}=-{{e}^{\nu(r)}}{{dt}^{2}}+{{e}^{\eta(r)}}{{dr}^{2}}+{r}^{2}({{d\theta}^{2}}+{{\sin}^{2}}\theta~{{d\phi}^{2}}).   
\end{equation}
where $\nu$ and $\eta$ are metric potentials and depend only on the
radial coordinate r, for which the staticity of spacetime is guaranteed. For a spherically symmetric stellar system, combining Eqs. (\ref{10}), (\ref{eq11}), (\ref{eq12}) and (\ref{eq14}), we arrive at the following differential equations
\begin{eqnarray}\label{eq15}
8\pi(\bar{\rho}+\alpha~\theta^{t}_{t})&=&{{\rm e}^{-\eta}}\left( {\frac {\eta^{{\prime}}}{r}}-\frac{1}{r^2}\right)+\frac{1}{r^2},~~~~~ \\ \label{eq16}
8\pi(\bar{p}-\alpha~\theta^{r}_{r})&=&{{\rm e}^{-\eta}} \left( {\frac {\nu^{{\prime}}}{r}}+\frac{1}{r^2}\right) -\frac{1}{r^2}, ~~~~~\\ \label{eq17}
8\pi(\bar{p}-\alpha~\theta^{\varphi}_{\varphi})&=&\frac{{\rm e}^{-\eta}}{4} \left(2 \nu^{{\prime\prime}}+\nu^{\prime 2}+2{\frac {\nu^{{\prime}}-\eta^{{\prime}}}{r}}-\nu^{\prime}\eta^{\prime} \right)
\end{eqnarray}
where
\begin{eqnarray}\label{eq18}
\bar{\rho}&=&\rho+\frac{\chi}{8\,\pi}(3\rho-p),\\\label{eq19}
\bar{p}&=&p -\frac{\chi}{8\,\pi}(\rho-3 p)
\end{eqnarray}
Then from the system (\ref{eq15})-(\ref{eq17}), we can write the corresponding conservation law $\nabla^{\mu}T^{(eff)}_{\mu\nu}=0$ as,
\begin{eqnarray}\label{eq20}
-\alpha\left[ \frac{\nu^{\prime}}{2}\left(\theta^t_t-\theta^r_r\right)-\frac{d \theta^r_r}{dr}+\frac{2}{r}\,(\theta^\varphi_\varphi-\theta^r_r)\right]-\frac{d\bar{p}}{dr}\nonumber\\-\frac{\nu^\prime}{2} (\bar{\rho}+\bar{p})=0.
\end{eqnarray}

As a whole the equations system (\ref{eq15})-(\ref{eq17}), we have three non-linear differential equations for seven unknown functions: $\{\eta, \nu\}$, $\{\rho^{(\text{eff})},
p^{(\text{eff})}\}$ and $\{\theta^{t}_{t}, \theta^{r}_{r}, \theta^{\varphi}_{\varphi}\}$, where these grouped functions in braces represent the metric potentials, the thermodynamic observables and the extra source components, respectively. So as to get these unknown we choose a systematic methodology. Moreover, the equations system (\ref{eq15})-(\ref{eq17}), which includes total energy density, total radial and total tangential pressures, can be classified as follows
\begin{eqnarray}\label{eq21}
\rho^{\rm{(eff)}}&=&\bar{\rho}+\alpha~\theta^{t}_{t}=\rho+\frac{\chi}{8\,\pi}(3\rho-p)+\alpha~\theta^{t}_{t}, \\ \label{eq22}
p^{(\text{eff})}_{r}&=&\bar{p}-\alpha~\theta^{r}_{r}=p -\frac{\chi}{8\,\pi}(\rho-3 p)-\alpha~\theta^{r}_{r}, \\ \label{eq23}
p^{(\text{eff})}_{t}&=&\bar{p}-\alpha~\theta^{\varphi}_{\varphi}=p -\frac{\chi}{8\,\pi}(\rho-3 p)-\alpha~\theta^{\varphi}_{\varphi}.
\end{eqnarray}
Once more, we remark that when $\theta^{r}_{r}\neq\theta^{\varphi}_{\varphi}$, anisotropic behavior appears in the system due to the present of the $\theta$-sector. So, we defined the effective anisotropy factor $\Delta^{(\text{eff})}$, in order to measure the anisotropy behavior as follows
\begin{equation}\label{eq24}
\Delta^{(\text{eff})}=p^{(\text{eff})}_{t}-p^{(\text{eff})}_{r}=\alpha\left(\theta^{r}_{r}-\theta^{\varphi}_{\varphi}\right).
\end{equation}
In this respect, the equations system (\ref{eq15})-(\ref{eq17}) could undoubtedly be dealt as an anisotropic fluid, which should make it possible to consider five unknown functions, viz., the two metric potentials and the three total functions expressed in the equations system (\ref{eq15})-(\ref{eq17}). However, we will perform an alternate methodology, as clarified below.

\section{Gravitational Decoupling by MGD Approach}\label{sec4}
In this part, we use the minimal geometric deformation gravitational decoupling procedure as previously mentioned. So, for extending spherical and static isotropic fluid solutions to anisotropic areas, it is necessary that the gravitational decoupling by MGD approach becomes a straightforward and efficacious implement. Let us consider the perfect fluid solution $\{\xi,\mu,\bar{\rho},\bar{p}\}$ by deactivating the coupling $\alpha$, so that the canonical line element (\ref{eq14}) becomes 
\begin{equation}\label{eq25}
{ds}^{2}=-{{e}^{\xi(r)}}{{dt}^{2}}+\frac{{dr}^{2}}{\mu(r)}+{r}^{2}({{d\theta}^{2}}+{{\sin}^{2}}\theta~{{d\phi}^{2}}),
\end{equation}
where $\mu(r)=1-2m/r$ is the standard term for the mass function. Now, in order to see the $\theta$-sector impacts generated by the gravitational source $\theta_{\mu\nu}$ in the equations system (\ref{eq15})-(\ref{eq17}), we have to incorporate the coupling parameter $\alpha$ in the perfect fluid distribution. In order to incorporate the impacts of the new sector $\theta_{\mu\nu}$ in uncharged anisotropic model, we consider the geometric deformation  functions ${\xi}$ and ${\mu}$ as
\begin{eqnarray}\label{eq26}
\xi\rightarrow \nu&=&\xi+\alpha h \\ \label{eq27}
\mu\rightarrow e^{-\eta}&=&\mu+\alpha f,
\end{eqnarray}
where $f$ and $h$ represent the deformations related to the components of the radial and temporal line elements, respectively. It should be noted that the previous deformations are purely radial functions, this characteristic ensures the spherical symmetry of the solution. The assumed MGD method relates to set $h=0$ or $f=0$, for this situation the deformation will be performed only on the radial component, remaining the temporal one unaltered, that is to say that $h=0$. In this respect where $h=0$, we have
\begin{equation}\label{eq28}
\mu(r)\rightarrow e^{-\eta(r)}=\mu(r)+\alpha f(r),    
\end{equation}
which is known as the Minimal Geometric Deformation (MGD).

Using now this Eq. (\ref{eq28}) is easy to see that the equations system (\ref{eq15})-(\ref{eq17}) splits in two equations systems. The first of them is
\begin{eqnarray}\label{eq29}
8\,\pi\,\bar{\rho}&=&-\frac{\mu^{\prime}}{r}-\frac{\mu}{r{^2}}+\frac{1}{r^{2}} =8\,\pi\rho+\chi\,(3\rho-p),
\end{eqnarray}
\begin{eqnarray}\label{eq30}
 8\,\pi\,\bar{p}&=&\mu \left( {\frac {{\nu}^{{\prime}}}{r}}+\frac{1}{r^2}\right) -\frac{1}{r^2} =8\,\pi\,p -\chi\,(\rho-3 p), 
 \end{eqnarray}
\begin{eqnarray}\label{eq31}
8\,\pi\,\bar{p}&=& \frac{\mu}{4} \left( {2\nu}^{{\prime\prime}}+{\nu}^{{\prime\,2}}+2\frac{{\nu}^{{\prime}}}{r}\right)+\frac{\mu^{\prime}}{4}\,\left( {\nu}^{{\prime}}+\frac{2}{r}  \right)\nonumber\\
&&\hspace{3cm}= 8\,\pi\,p -\chi\,(\rho-3 p),~~~~~
\end{eqnarray}
which corresponds to $\alpha= 0$, implying that the distribution of the fluid matter is perfect.

From now we will call the  equations system (\ref{eq29})-(\ref{eq31}) as modified $f(R, T)$ gravity system. Furthermore, with the considerations relating to Eqs.(\ref{eq18}) and (\ref{eq19}), the quantities $\rho$ and $p$ can be written in terms of metric potentials as follows
\begin{eqnarray}
\rho=\frac{1}{(8\pi+4\chi)\,(8\pi+2\chi)\,r^2}\,\big\{(8\pi+3\chi)(1-\mu^{\prime}\,r-\mu)\nonumber\\+\chi\,(\mu\,\nu^{\prime}\,r+\mu-1)\,\big\},~~~~~\label{eq32}\\
p=\frac{1}{(8\pi+4\chi)\,(8\pi+2\chi)\,r^2}\,\big\{(8\pi+3\chi)\,(\mu\,\nu^{\prime}\,r+\mu-1)\nonumber\\+\chi\,(1-\mu^{\prime}\,r-\mu)\,\big\}.~~\label{eq33}~~
\end{eqnarray}
while the pressure isotropy equation $G^1_1=G^2_2$ reduces to
\begin{equation}\label{eq34}
\begin{split}
4\left(1-\mu\right)+2r\left(\mu^{\prime}-\mu\nu^{\prime}\right)+r^{2}\left(2\mu\nu^{\prime\prime}+\mu\nu^{\prime 2}+\mu^{\prime}\nu^{\prime}\right)=0.   
\end{split}
\end{equation}
So from this Eq. (\ref{eq34}) we notice that all the solutions obtained in the general relativity case are also present in the modified $f(R, T)$ gravity theory case. Moreover, it is clear that, despite the qualitative subtlety between the solutions for Einstein's theory of gravity and the modified $f(R, T)$ theory of gravity, there is a quantitative difference between the two models when considering the geometric and material content. In this regard the two models share only the geometrical content but not the material one, is in this situation that any solution to Einstein theory of gravity can be viewed as a solution in the gravitational $f(R, T)$ system. The standard expression for the mass takes the new form as,
\begin{eqnarray}
m(r)&=&4\pi\int_0^r{\rho(r)\, r^2\,dr}\label{eq35}\\
m(r)&=&\frac{4\pi}{[(8\pi+3\chi)^2-\chi^2]}\,\int^r_0 \Big\{2\,(4\pi+\chi)\,(1-e^{-\lambda})\nonumber\\&+&(8\pi+3\chi)\lambda^{\prime}\,r\,e^{-\lambda}+\chi\,\nu^{\prime}\,r\,e^{-\lambda} \Big\}\,dr,\label{eq36} 
\end{eqnarray}

Here $\chi$ is a dimensionless coupling constant, which plays a fundamental role for determining the astrophysical configuration. At this regard, it is essential to point out that when the coupling constant $\chi$ is zero,  the modified $f(R, T)$ gravity recovers the same physical behavior as the general relativity. It should be noted that we have considered the Eq. (\ref{eq34}), as the main differential equation in this study contains two fundamental integration constants. In any case, we can adapt the constants regarding the mass $M$ and radius $R$ of the distribution by solving the linear corresponding equation.

As previously mentioned, the quantities $\rho$ and $p$ after some algebraic handling in their proper formulas contain the coupling constant as anticipated. Moreover, using the equations (\ref{eq32})-(\ref{eq33}) with respect to $\chi=0$ we find the standard GR field equations for distributions of the isotropic matter. additionally, combining Eqs (\ref{eq32}) and (\ref{eq33}), we obtain the usual inertial mass density $\rho+p$ in the following form
\begin{equation}\label{eq37}
\rho+p=\frac{\mu \nu^{\prime}-\mu^{\prime}}{r}.
\end{equation}
On the other hand, the second equations system reads
\begin{eqnarray}\label{eq38}
 -\frac{f{^\prime}}{r}-\frac{f}{r^2}&=& 8\pi \theta^{t}_{t} , \\ \label{eq39}
-f \left( {\frac {{\nu}^{{\prime}}}{r}}+\frac{1}{r^2}\right) &=& 8\pi \theta^{r}_{r}  \\ \label{eq40}
-\frac{f}{4} \left( {2\nu}^{{\prime\prime}}+{\nu}^{{\prime\,2}}+2\frac{{\nu}^{{\prime}}}{r}\right)-\frac{f^{\prime}}{4}\left( {\nu}^{{\prime}}+\frac{2}{r}  \right)&=& 8\pi \theta^{\varphi}_{\varphi},
\end{eqnarray}

and the conservation equations associated with the equations ensemble (\ref{eq32})-(\ref{eq33}) and (\ref{eq36})-(\ref{eq38}) are

\begin{eqnarray}\label{eq41}
\frac{{\nu^\prime}}{2}\,({\rho}+{p})+\frac{d{p}}{dr}-{\frac {\chi}{2(4\pi+\chi)}}\frac{d}{dr}\left(\rho-p\right)=0, \\ \label{eq42}
-\frac{{\nu^\prime}}{2}\,(\theta^t_t-\theta^r_r)+\frac{d \theta^r_r}{dr}-\frac{2}{r}\,(\theta^\varphi_\varphi-\theta^r_r)=0.
\end{eqnarray}
 
Furthermore, we remark that the linear combination of conservation equations (\ref{eq41}) and (\ref{eq42}) with respect to coupling constant $\alpha$ gives the conservation equation for the total stress-energy tensor ${T}^{\mu{(\text{eff})}}_{\nu}={\bar{T}}^{\mu}_{\nu} + \alpha\theta^{\mu}_{\nu}$, in the form
\begin{eqnarray}\label{43}
&& -\frac{d{p}}{dr}-\alpha\left[ \frac{{\nu^\prime}}{2}\,(\theta^t_t-\theta^r_r)-\frac{d \theta^r_r}{dr}+\frac{2}{r}\,(\theta^\varphi_\varphi-\theta^r_r)\right]-\frac{{\nu^\prime}}{2}\,({\rho}+{p}) \nonumber \\
&& \hspace{3.9cm} +{\frac {\chi}{2(4\pi+\chi)}}\frac{d}{dr}\left(\rho-p\right)=0.
\end{eqnarray}

So from this Eq. (\ref{43}), we notice that the last factor is an additional term which is coupled via $\chi$ parameter in the $f(R, T)$ gravity arena. Furthermore, this additional term could on a fundamental level be attractive or repulsive in nature, due to its dynamical behavior of the physical parameters like energy density and pressure depends on the behavior of the free parameter $\chi$. In the limiting case of $\chi=0$, Eq. (\ref{43}) is similar to the explicit form of conservation law $\nabla^{\mu}T^{(eff)}_{\mu\nu}=0$ given in Eq. (\ref{eq20}).\\

Now it is important to remark that from now onward we will characterize the complete physical parameters for the energy density, radial pressure and tangential pressure as follows :
\begin{eqnarray}
\rho^{\text(tot)}(r)&=&\rho(r)+\alpha~\theta^{t}_{t}(r), \label{45}\\
p_{r}^{\text(tot)}(r)&=&p(r)-\alpha~\theta^{r}_{r}(r), \label{eq46}\\
p_{t}^{\text(tot)}(r)&=&p(r)-\alpha~\theta^{\varphi}_{\varphi}(r), \label{eq47}
\end{eqnarray}
where the quantities $\rho$ and $p$ are specified by Eqs. (\ref{eq32}) and (\ref{eq33}), respectively. On the other hand, we can see that the Eqs. (\ref{eq32})-(\ref{eq33}) depends on the extra geometric term which is coupled by means of $\chi$ parameter. For this purpose, we can conclude that the physical quantities such as, the total anisotropy factor ($\Delta^{\text{tot}}$) and the effective anisotropy factor ($\Delta^{\text{eff}}$) are characterized by the Eqs. (\ref{eq46})-(\ref{eq47}) and Eq. (\ref{eq24}) respectively, are indistinguishable. Now the inner stellar geometry for the present MGD model can be given by the following line element, 
\begin{equation}\label{eq40aa}
ds^{2}=-{{e}^{\nu(r)}}{{dt}^{2}}+\left[1-\frac{2\,\bar{m}(r)}{r}\right]^{-1}{{dr}^{2}}+{r}^{2}({{d\theta}^{2}}+{{\sin}^{2}}\theta{{d\phi}^{2}}),
\end{equation}
where $\bar{m}(r)$ defines the internal mass of the MGD model which can be given as,  
\begin{equation}\label{eq41aa}
\bar{m}(r)=\frac{r}{2}\left[\frac{2\,m(r)}{r}-{\alpha}\,f(r)\right].    
\end{equation}

\section{Exterior space-time: Junction conditions}\label{sec5}
To ensure a well behaved compact structure i.e., a confined and finite matter distribution with well posed mass $M$ and radius $R$, it is necessary to join the inner geometry $\mathcal{M^{-}}$ at the surface $\Sigma=r=R$ with the exterior space-time $\mathcal{M^{+}}$ surrounding the configuration. In this respect, in the case of general relativity framework the outer manifold is well known, precisely it corresponds to Schwarzschild vacuum space-time (in considering uncharged, non-radiating and static compact object). However, in the context of $f(R,{T})$ gravity the outer manifold surrounding the fluid sphere  does not necessarily coincide with the Schwarzschild solution, what is more this exterior space-time could in principle receive contributions from the material sector given by the trace of the energy-momentum tensor due to the breakdown of the minimum coupling matter principle between the gravitational  and the material sector. Thus, the usual junction conditions applicable in general relativity i.e., Israel-Darmois (ID from now on) \cite{Israel:1966a,Darmois:1927a} matching conditions could not work in this context any more or should be appropriate redefined. So, a simple way to see how the contributions coming from the $f(R,{T})$ function could affect the exterior space-time, is re-written the field equations (\ref{6}) as follows (putting $\alpha=0$),
\begin{equation}\label{eqq48}
\begin{split}
G_{\mu\nu}\equiv R_{\mu\nu}-\frac{R}{2}g_{\mu\nu}=\frac{1}{f_{R}}\bigg[8\pi T_{\mu\nu}+\frac{f}{2}g_{\mu\nu} -\frac{R}{2}f_{R}~ g_{\mu\nu} &\\
-\left(T_{\mu\nu}+\Theta_{\mu\nu}\right)f_{T}-\left(g_{\mu\nu}\Box-\nabla_{\mu}\nabla_{\nu}\right)f_{R}\bigg],   
\end{split}    
\end{equation}
then by taking the trace of the above equation one obtains
\begin{equation}\label{eqq49}
\begin{split}
R=\frac{1}{f_{R}}\bigg[8\pi T+2f-\left(T+\Theta\right)f_{T}-3\, \Box f_{R}\bigg].
\end{split}
\end{equation}
So, in considering an empty matter field i.e, $T_{\mu\nu}=0 \rightarrow T=0$ Eq. (\ref{eqq49}) yields to
\begin{equation}\label{eqq50}
R=\frac{1}{f_{R}}\left[2f_{1}-3\, \Box f_{R}\right],   
\end{equation}
where $f_{1}$ represents the geometrical part of the $f(R,{T})$ function that is $f_{1}\equiv f_{1}(R)$. This is so because the $f(R,{T})$ function can be seen as $f(R,{T})=f_{1}(R)+f_{2}(T)$. Therefore, it is clear that a vanishing energy-momentum tensor in the framework of $f(R,{T})$ gravity theory does not mean a null Ricci scalar like in general relativity where for $T_{\mu\nu}=0\rightarrow R=0 \rightarrow R_{\mu\nu}=0$ which represents a vacuum space-time. What is more, $T_{\mu\nu}=0$ does not imply $f_{2}=0$ at all, of course this term could contributes with a constant term for example. However, if $T_{\mu\nu}=0$ leads to $f_{2}=0$, the outer manifold is affected by the geometric terms encoded in $f_{1}$ and $f_{R}$. This last situation also happens in $f(R)$ gravity theory, where $R^{2}$, $R^{3}$ and so on,  modify the interface between the inner geometry and the exterior one \cite{Capozziello:2002b}.

At this point it is clear that the ID conditions will work as they do in general relativity, if and only if one ensures that the external solution matches those of general relativity. Obviously it is not an easy task to achieve it, since the function $f(R,T)$ can be as complex as one wants. As regards this, equation (\ref{eqq12}) ensures that the contributions of the geometric and material sector remain confined within the range $0\leq r\leq R$. Moreover, the fact that the function $f(R,T)$ is linear in $R$ plus a linear coupling in $T$ through an adjustable parameter $\chi$, can be seen as a general relativity model coupled to a variable cosmological constant which also breaks the minimal coupling matter with the gravitational sector. Concretely, by inserting (\ref{eqq12}) into Eq. (\ref{eqq50}) and solving for $R$ provides $R=0$ and hence $R_{\mu\nu}=0$ (in Eq. (\ref{eqq48})) which represents a vacuum exterior space-time \i.e, the Schwarzschild geometry surrounding the fluid sphere. On the other hand, only the potential contributions coming from the $f(R,T)$ model was analyzed, however one needs to study the behaviour of the extra piece of the energy-momentum tensor i.e., the $\theta$-sector. This new term could also, in principle, modify the material content of the external space-time. So, the specific geometry describing this external manifold will be given by
\begin{equation}\label{eqq51}
ds^{2}=-\left[1-\frac{2{M}}{r}\right]dt^{2}+ \frac{dr^2}{ 1-\frac{2{M}}{r}+\alpha\, g(r)} +r^{2}d\Omega^{2},
\end{equation}
where $g(r)$ is the  geometric deformation for the exterior Schwarzschild space-time due to  $\theta_{\mu\nu}$ source. The metric (\ref{eqq51}) represents a deformed Schwarzschild space-time which is not vacuum any more. However, without loss of generality and for the sake of simplicity one can render $g(r)$ to be null recovering the usual outer vacuum space-time. 

So, to join in a smoothly way the internal configuration with the exterior one, the ID junction conditions involve the first and second fundamental forms. The first fundamental form express the continuity of the metric potentials across the boundary $\Sigma$. More specifically, the metric potentials describe the intrinsic geometry of the manifolds. So, the first fundamental form reads
\begin{equation}\label{eq52a}
\left[ds^{2}\right]_{\Sigma}=0,
\end{equation}
concisely
\begin{equation}\label{eq48}
e^{\nu^{-}(r)}|_{r=R}=e^{\nu^{+}(r)}|_{r=R},
\end{equation}
and
\begin{equation}\label{eq49}
e^{\eta^{-}(r)}|_{r=R}=e^{\eta^{+}(r)}|_{r=R},
\end{equation}
standing $"-"$ and $"+"$ the inner and external geometry respectively. The second fundamental form is related with the continuity of the extrinsic curvature $K_{\mu\nu}$ induced by $\mathcal{M}^{-}$ and $\mathcal{M}^{+}$ on $\Sigma$. The continuity of $K_{rr}$ component across $\Sigma$ yields to
\begin{equation}\label{eq50}
\left[p^{(\text{tot})}_{r}(r)\right]_{\Sigma}=\left[p(r)-\alpha\, \theta^{r}_{r}(r)\right]_{\Sigma}=0.
\end{equation} 
At this stage we would like to mention that, $p^{(\text{tot})}_{r}$ has taken a different form in comparison with the expression (\ref{eq22}) since $\rho$ and $p$ were separated implying that the coupling parameter $\chi$ is involved.  Now the terms coming from the coupling parameter are encoded in separate expressions for $\rho$ and $p$ given by Eqs. (\ref{eq32})-(\ref{eq33}). From this point of view it is clear how $\chi$ contribution comes into the pressure equation. Now we shall denote the extra components for coupling parameter $\chi$ in the expression (\ref{eq33}) as follows
\begin{equation}\label{eq58}
F_{\chi}(r)=\chi\,F^1_{\mu\nu}(r)-h_{\chi}\,\sum_{n=0}^{\infty}(-1)^{n}(h_{\chi})^{n}\,[F^2_{\mu\nu}(r)+\chi\,F^1_{\mu\nu}(r)].
\end{equation}
where, 
\begin{eqnarray}
&&F^1_{\mu\nu}(r)=\frac{3\mu}{8\pi} \bigg(\frac{\nu^{\prime}}{r}+\frac{1}{r^2} \bigg)-\frac{1}{8\pi} \bigg(\frac{\mu^{\prime}}{r}+\frac{\mu}{r^2}-\frac{1}{r^2}\bigg),\nonumber\\
&&F^2_{\mu\nu}(r)= -\bigg(\frac{\mu^{\prime}}{r}+\frac{\mu}{r^2}\bigg)+\frac{1}{r^2},~~~~h_{\chi}=\bigg(\frac{\chi}{4\pi}+\frac{\chi^2}{8\pi^2}\bigg).\nonumber
\end{eqnarray}
It should be noted that the form of $F_{\chi}$ depends on the choice of $T_{\mu\nu}$ which in our case is given by Eq. (\ref{10}) describing a perfect fluid matter distribution. At this point a couple of comments are pertinent. First, $p(r)$ in Eq.(\ref{eq46}) is given by Eq. (\ref{eq33}). 
Second, in this way the $f(R,T)$ contribution will come into the $\theta$-sector through the decoupler function $f(r)$ (as we will see later) in order to see the effects on it. So Eq. (\ref{eq50}) reads
\begin{equation}\label{eq52}
\left[p(r)-\alpha\, \theta^{r}_{r}(r)\right]_{r=R^{-}}=
\left[-F_{\chi}(r)-\alpha\, \theta^{r}_{r}(r)\right]_{r=R^{+}}.
\end{equation}
 Taking into account the previous discussion Eq. (\ref{eq52}) becomes to 
\begin{equation}\label{eq54}
\left[p(r)-\alpha\, \theta^{r}_{r}(r)\right]_{r=R^{-}}=
0.
\end{equation}
Moreover, by using Eq. (\ref{eq39}) in (\ref{eq50}) we obtain 
\begin{equation}\label{eq57}
\begin{split}
p(R)+\alpha f(R)\left(\frac{1}{R^{2}}+\frac{\nu^{\prime}(R)}{R}\right)=0.
\end{split}
\end{equation}
As said earlier, equation (\ref{eq57}) is an important result since the compact object will therefore be in equilibrium in a true exterior space-time without material content (vacuum) only if the total radial pressure at the surface vanishes. The obtained condition (\ref{eq57}) determines the size of the object \i.e the radius $R$ which means that the material content is confined within the region $0\leq r\leq R$. Furthermore, the continuity of the remaining components $K_{\theta\theta}$ and $K_{\phi\phi}$ leads to
\begin{equation}\label{eq59}
\bar{m}(R)=\bar{M}.
\end{equation}
Equation (\ref{eq59}) refers to the total effective mass contained in the sphere.

\section{Mimic constant}\label{sec6}

It is interesting to consider the form adopted by the field equations (\ref{eq32})-(\ref{eq33}) and (\ref{eq38})-(\ref{eq40}). In this respect, we incorporate the gravitational decoupling technique (\ref{eq28}), one obtain a few limitations imposed on these field equations, namely the mimic constraints. Furthermore, these options lead to new solutions are well behaved from the physical and mathematical point of view as well as free of geometrical singularities,  violation of the causality condition, non-decreasing thermodynamic functions etc. Moreover, other choices can be fulfilled through some basic requirements in order to be an admissible model from the physical and mathematical point of view. In this regard, a direct and satisfactory portrayal for the geometric deformation function $f(r)$ \cite{Morales:2018aa,Morales:2018a,Estrada:2019a,Maurya:2019a} should be fulfilled with these basic requirements or only concern the components of $\theta$-term with a barotropic, polytropic or linear state equation.

\subsection{$\theta$-effects: Mimic constraint on density for anisotropy}\label{B}
 The necessary and sufficient conditions for a smooth admissible interior solution in the interior spherical configuration is given when constraining the related radial pressure $\theta^{r}_{r}$ to mimic the density $\rho(r)$. Expressly
\begin{equation}\label{eq60}
\theta^{t}_{t}(r) =\rho(r).  
\end{equation}
Hence, by comparing with Eqs. (\ref{eq32}) and (\ref{eq38}), then the decoupler function $f(r)$ can be given as, 
\begin{eqnarray}\label{eq61}
f(r)=\frac{8\pi}{(8\pi+4\chi)(8\pi+2\chi)}\,\bigg((8\pi+4\chi)\,\mu(r)-(8\pi+2\chi)\nonumber\\-\frac{\chi}{r}\int{\mu(\nu^{\prime}r+1)\,dr}\bigg).~~~~~~~
\end{eqnarray}

Thus the deformed gravitational potential $e^{-\eta}$ associated to radial metric component $\mu$ can given through the eq.(\ref{eq28}) as, 
\begin{eqnarray}\label{eq62}
e^{-\eta}=\mu+\frac{8\pi}{(8\pi+4\chi)(8\pi+2\chi)}\,\bigg[(8\pi+4\chi)\,\mu(r)-\nonumber\\
(8\pi+2\chi)-\frac{\chi}{r}\int{\mu(\nu^{\prime}r+1)\,dr}\bigg].~~~~~~~   
\end{eqnarray}
Then we can acquire, 
\begin{eqnarray}\label{eq63}
p_{r}^{(\text{tot})}&=&p(r)-\alpha\, \theta^{r}_{r} \\ \label{eq64}
p_{t}^{(\text{tot})}&=&p(r)-\alpha\, \theta^{\varphi}_{\varphi},
\end{eqnarray}
and making use of the formula ($\rho^{(\text{tot})}=\rho+\alpha\, \theta^{t}_{t}$), with the help of condition (\ref{eq60}), we can express the $\rho^{(\text{tot})}$ as, 
\begin{equation}\label{eq65}
\rho^{(\text{tot})}=\left(1+\alpha\right)\rho(r),  
\end{equation}
From Eq.(\ref{eq62}), we note that the gravitational potential $\eta$ contains integral terms therefore it is not always easy to obtain deformed radial gravitational potential $e^{\eta}$ for a given function of $\nu$. Moreover, from Eqs. (\ref{eq32}), (\ref{eq33}) and (\ref{eq62}) one can obtain the energy density $\rho(r)$ and the pressure $p(r)$ for a specific metric potential $\nu$. 

\subsection{$\theta$-effects: Mimic constraint on pressure for anisotropy}\label{A}
In this situation, we apply another approach of isotropic pressure given by Eq. (\ref{eq33}) mimics its analogy of the anisotropic sector indicate in Eq. (\ref{eq39}) to close equations system (\ref{eq38})-(\ref{eq40}) in order to obtain a physically and mathematically acceptable solution. The simplest approach which fulfills this crucial requirement is

\begin{equation}\label{eq66}
\theta^{r}_{r}(r)=p(r). 
\end{equation}
This requirement infers that the energy-momentum tensor for the seed solution matches with the anisotropy in the radial direction. As a result, the fact that Eq. (\ref{eq33}) and (\ref{eq39}) are equal, which gives us a general form of the generic function which involves deformation function as
\begin{eqnarray}\label{67}
f(r)=\frac{(8\pi+3\chi)\,(1-\mu\,\nu^{\prime}\,r-\mu)-\chi\,(1-\mu^{\prime}\,r-\mu)}{(8\pi+4\chi)\,(8\pi+2\chi)\,(\nu^{\prime}\,r+1)/8\pi}.~~~~~~
\end{eqnarray}

Then, the general form of minimally deformed radial gravitational potential $e^{-\eta}$ is explicit as
\begin{eqnarray}\label{68}
e^{-\eta}=\mu+\frac{(8\pi+3\chi)\,(1-\mu\,\nu^{\prime}\,r-\mu)-\chi\,(1-\mu^{\prime}\,r-\mu)}{(8\pi+4\chi)\,(8\pi+2\chi)\,(\nu^{\prime}\,r+1)/8\pi\alpha}.\nonumber\\   
\end{eqnarray}
In this scenario we can obtain,  
\begin{eqnarray}
p_{r}^{(\text{tot})}&=& (1-\alpha)\,p(r)  \label{eq72}
\end{eqnarray}
\begin{eqnarray}
p_{t}^{(\text{tot})}&=&p_{r}^{(\text{tot})}+\Delta, \label{eq73}
\end{eqnarray}
and by virtue of condition (\ref{eq66}) we can write the explicit form of $\rho^{(\text{tot})}$ as,
\begin{equation}\label{eq74}
\rho^{(\text{tot})}=\rho(r)+\alpha\,\theta^{\varphi}_{\varphi},  
\end{equation}
Here we see that  Eq.(\ref{68}) does not contains any integral term. This implies that the deformed metric potential $e^{\eta}$ can be obtained easily in the case of $\theta^r_r=p(r)$ as compared to $\theta^t_t=\rho(r)$, where the energy density $\rho(r)$ and the pressure $p(r)$ are given by Eqs. (\ref{eq32}) and (\ref{eq33}), respectively.

\section{DECOUPLING MGD STELLAR STRUCTURE : Korkina-Orlyanskii  model }\label{sec7} 
In the present section we wish to obtain a gravitational decoupling MGD model in  the framework of modified $f(R,T)$ gravity theory. For this purpose we will solve the system (\ref{eq38})-(\ref{eq40}) by employing an appropriate restriction on components of $\theta_{\mu\nu}$ to determine the deformation function $f(r)$. But, in order to determine the deformation function $f(r)$ it is necessary to specify the seed solution which must satisfy the eq. (\ref{eq34}). Therefore, we take Korika-Orlyanskii perfect fluid solution as a seed solution which is satisfying the isotropic condition (\ref{eq34}). In doing so, we follow the same techniques as adopted by Ovalle to close the system (\ref{eq38})-(\ref{eq40}). The gravitational potentials corresponding to Korika-Orlyanskii solution are given as,    
\begin{eqnarray}
 &&\mu=\frac{1+Ar^2}{1+2Ar^2},  \label{72}\\
&&e^{\nu}=C\,(1+Ar^2)\,\bigg[D+\frac{\sqrt{1+2Ar^2}}{1+Ar^2}\nonumber\\&& \hspace{0.8cm}-\sqrt{2}\, \label{73} \ln\big(\sqrt{1+2Ar^2}+\sqrt{2+2AR^2}\big) \,\bigg]^2. \label{hyper}
\end{eqnarray}
where $C$ and $D$ are the arbitrary integration constants. Note that the ansatz for the metric function $g_{rr}$ in Eq. (\ref{72}) firstly suggested by Korikina and Orlyanskii \cite{Korkina:1991a} to develop a method for generating spherically symmetric static solutions of the field equations, and consequently used by Maurya and Gupta \cite{Maurya:2012a} to build the well-behaved relativistic charged compact spherical object systems as well as the role of pressure anisotropy on the same objects which is developed in \cite{Maurya:2018aa}. This choice of the metric function is physically well-motivated and has been utilized by several researchers in the past to build feasible astrophysical systems. Moreover, the metric function (\ref{72}) is likewise positive and free from the geometric singularity at the center. 
The density and isotropic pressure for Korika-Orlyanskii model in the framework $f(R,T)$ gravity theory can be expressed as
\begin{eqnarray}
 &&\hspace{-0.15cm} \rho =  \frac{A}{(64\pi^2+48\pi\chi+8\chi^2)}\bigg[\frac{(8\pi+3\chi)\,(3+2Ar^2)}{(1+2Ar^2)^2}\nonumber\\&&\hspace{0.25cm} + \frac{\chi}{(1+2Ar^2)}-\frac{4\chi\,[1 + A^2r^4 + Ar^2\,(2 +\sqrt{1 + Ar^2})]}{(1 + 3Ar^2+2A^2r^4)^{3/2}\,[D+\Psi(r)]}\bigg],~~~~~~\label{eqda}\\
&&\hspace{-0.15cm}p =  \frac{A}{(64\pi^2+48\pi\chi+8\chi^2)}\bigg[\frac{\chi\,(3+2Ar^2)}{(1+2Ar^2)^2}+ \frac{(8\pi+3\chi)}{(1+2Ar^2)}\nonumber\\&&\hspace{0.9cm}-\frac{4\,(8\pi+3\chi)\,[1 + A^2r^4 + Ar^2\,(2 +\sqrt{1 + Ar^2})]}{(1 + 3Ar^2+2A^2r^4)^{3/2}\,[D+\Psi(r)]}\bigg].~~ \label{eqpa}
\end{eqnarray}
where,\\ $\Psi(r)=\frac{\sqrt{1+2Ar^2}}{1+Ar^2}- \sqrt{2}\, \ln\big(\sqrt{1+2Ar^2}+\sqrt{2+2AR^2}\big)$.\\
It is found that the Eqs. (\ref{eqda})-(\ref{eqpa}) consist of coupling parameter $\chi$. As a result, for $\chi\rightarrow 0$, the expressions (\ref{eqda})-(\ref{eqpa}) seems promising to explain the corresponding energy-density and the isotropic pressure of the original Korika-Orlyanskii, similar to energy-density and the isotropic pressure, which fulfilled the Einstein field equations. On the other hand, the expressions (\ref{eqda})-(\ref{eqpa}) shows that all the solutions obtained in the case of any perfect fluid solution of Einstein theory are also present in the context of modified $f(R, T)$ gravity theory due to the non-minimal coupling matter presented by the coupling parameter $\chi$, which gives a more complicated thermodynamic behavior.
In order to depict the expressions for $\theta-$components of the stellar models in $f(R, T)$ gravity we need to determine the deformation function $f(r)$. On the other hand, as we can see that the field Eqs. (\ref{eq32})-(\ref{eq33}) depends on metric function $e^{\nu}$ corresponding to Korika-Orlyanskii solution that involves the logarithmic relationship. In this situation, it is difficult to integrate the deformation function $f(r)$ in closed form in case of mimic constraint on density for anisotropy viz. $\theta^{t}_{t}(r) =\rho(r)$.  Therefore, to find the deformation function $f(r)$, we take the pressure of the anisotropy as mentioned above by eq.(\ref{eq66}) i.e. $\theta^r_r=p(r)$. Thus, by employing eqs.(\ref{67}), (\ref{72}) and (\ref{73}), we can obtain $f(r)$ in the following form,
\begin{eqnarray}
f(r)=\frac{ -\pi\,\chi\,(3 + 2 A r^2)\,A r^2\, +\pi\, (3 \chi + 8 \pi)  \,\Psi_2(r)}{(\chi+ 2 \pi)\,(\chi + 4 \pi) (1 + 2 A r^2)^2 \,[1 + 2 A r^2 \Psi_1(r)]},~~~~~
\end{eqnarray}
where,

\begin{widetext}

\begin{eqnarray}
\Psi_2(r) &=& (1 + 2 A r^2)\,A r^2\,[1-2 (1 + A r^2)\,\Psi_1(r)]~~,~~~
\Psi_1(r) = \frac{1}{(1 + A r^2)}-\frac{2\,[1 + A^2 r^4 + A r^2\,(2 +\sqrt{1 + A r^2})]}{[D + \Psi(r)]\, (1 + A r^2)^{5/2}\,\sqrt{1 + 2 A r^2}}.\nonumber
\end{eqnarray}

Hence the deformed gravitational potential $e^{-\eta(r)}$ turns out to be,
{\small
\begin{eqnarray}
e^{-\eta(r)}=\mu+\alpha\,f(r)=\frac{1+Ar^2}{1+2Ar^2}+\alpha \left(\frac{ -\pi\,\chi\,(3 + 2 A r^2)\,A r^2\, +\pi\, (3 \chi + 8 \pi) (1 + 2 A r^2)\,A r^2\, \,[1-2 (1 + A r^2)\,\Psi_1(r)]}{(\chi+ 2 \pi)\,(\chi + 4 \pi) (1 + 2 A r^2)^2 \,[1 + 2 A r^2 \Psi_1(r)]}\,\right).\label{eq80} 
\end{eqnarray}
}
So using Eqs. (\ref{eq72})-(\ref{eq74}) and (\ref{eq80}) , we get total pressures and total energy density as,
{\small
\begin{eqnarray}
&&p^{tot}_r=(1-\alpha)\,p=\frac{A\,(1-\alpha)}{(64\pi^2+48\pi\chi+8\chi^2)}\bigg[\frac{\chi\,(3+2Ar^2)}{(1+2Ar^2)^2}+ \frac{(8\pi+3\chi)}{(1+2Ar^2)}-\frac{4\,(8\pi+3\chi)\,[1 + A^2r^4 + Ar^2\,(2 +\sqrt{1 + Ar^2})]}{(1 + 3Ar^2+2A^2r^4)^{3/2}\,[D+f(r)]}\bigg],~~~~~~\\
&& p^{tot}_t=p^{tot}_r+\Delta, \\
&& \rho^{tot}=\rho +\alpha\,\theta^t_t
\end{eqnarray}}
\end{widetext}

where pressure anisotropy $\Delta=\theta^{\varphi}_{\varphi}-\theta^r_r$. 
From the above thermodynamics observable, we can determine that how gravitational decoupling by means of MGD effects on Korika-Orlyanskii model in the framework of modified gravity theory. In order to compare and discuss the results, we have taken four cases namely GR, GR+MGD, $f\left(R,T\right)$ and $f\left(R,T\right)$+MGD.   

Presently we have in particular the constants $C$ and $D$. These are characterizing inside the solution that can be obtain from Eqs. (\ref{eq48}), (\ref{eq49}) and (\ref{eq54}). The eqs. (\ref{eq48}) and (\ref{eq49}) leads to 
\begin{equation}
e^{\nu(r)}|_{r=R^{-}}=\left[\mu(r)+\alpha f(r)\right]|_{r=R^{-}}=1-2\frac{{M}}{R^{+}},
\end{equation}
where the Schwarzschild mass ${M}$ matches with all-out mass $\bar{M}$ contained in the spherical object at the limit $\Sigma$. Moreover the condition (\ref{eq54}) together with eq.(\ref{eq72}) on boundary $\Sigma$ yields,
\begin{equation}\label{82}
\left(1-\alpha\right)p(r)|_{r=R^{-}}=0.
\end{equation}
We find from the above Eq.(\ref{82}) the constant $D$ as,
\begin{eqnarray}
D=\frac{\sqrt{1 + 2 AR^2}~ [4\pi\,D_1(R) + \chi D_2(R)] +  D_3(R)}{(1 + AR^2)^{3/2}\, (3 \chi + 4 \pi + 4 \chi AR^2 + 8 \pi AR^2)}.~~~
\end{eqnarray}
where, 
{\small
\begin{eqnarray}
D_1(R)&=&4 + 4 A^2R^4 - \sqrt{1 + AR^2} + 2AR^2 (4 + \sqrt{1 + AR^2}),\nonumber\\
D_2(R)&=&6 + 6 A^2R^4 - 3 \sqrt{1 + AR^2} + 2 AR^2 (6 + \sqrt{1 + AR^2}),\nonumber\\
D_3(R)&=&(3 \chi + 4 \pi + 4 \chi AR^2 + 8 \pi AR^2)(1 + AR^2)^{3/2}\nonumber\\&&\sqrt{2}\,\ln\big[\sqrt{1 + 2 AR^2} + \sqrt{2 + 2 AR^2}\big].\nonumber
\end{eqnarray}}
However, the constant $C$ can be determined by relation $e^{\nu(r)}|_{r=R^{-}}=\left[\mu(r)+\alpha f(r)\right]|_{r=R^{-}}$ which is needed in order to determine the gravitational redshift within the MGD model.  
Consequently, the natural constraint on the free parameter $\alpha$ is obtained from the last Eq. (\ref{82}) (since pressure cannot be negative) as 
\begin{equation}\label{eq67}
\alpha<1,    
\end{equation}
In order to conserve $p^{(\text{tot})}_{t}>p^{(\text{tot})}_{r}$ at all focuses interior the collapsed configuration which guarantees also $\Delta>0$, what keeps the framework from carrying out unwanted actions such as instabilities.
As we can see, $p^{(\text{tot})}_{t}$ imposes a lower limit on $\alpha$ \i.e, $\alpha>0$. Thus, the positivity of the complete tangential pressure in the whole compact structure is guaranteed. So we have
\begin{equation}\label{eq88}
0<\alpha<1,    
\end{equation}

\section{Dynamical Equilibrium Condition}\label{sec8}
\begin{figure}[!t]
\includegraphics[width=8cm,height=6cm]{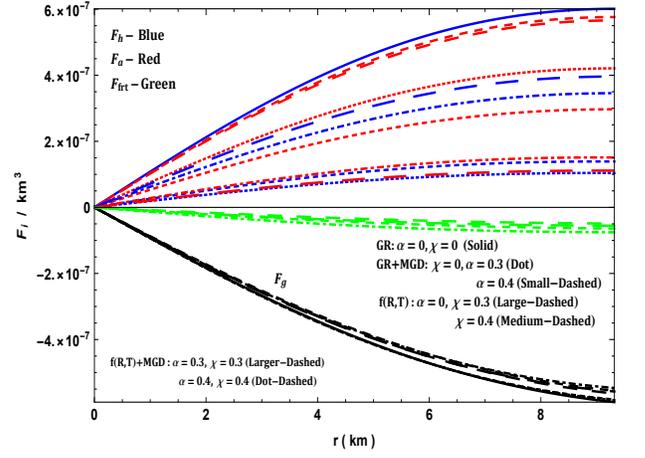}
\caption{Variation of different gradients in TOV-equation with radial coordinate $r$. where we define $F_h=-\frac{dp}{dr}-\alpha\,\frac{d\theta^r_r}{dr}$, $F_a=\frac{2\,\alpha}{r}\,(\theta^r_r-\theta^\varphi_\varphi)$, $F_g=-\frac{{\nu^\prime}}{2}\,({\rho}+{p})-\frac{{\alpha\,\nu^\prime}}{2}\,(\theta^t_t-\theta^r_r)$, and $F_{frt}={\frac {\chi}{2(4\pi+\chi)}}\frac{d}{dr}\left(\rho-p\right)$. }\label{f12}
\end{figure}

In order to perform the equilibrium analysis of the model for stellar system through MGD method in the context of $f(R,T)$ gravity theory, we need the generalized TOV equation. It comes from the modified form of the energy conservation equation for the stress-energy tensor in the same modified gravity framework given in Eq. (\ref{eq13}). In this respect, the modified form of the TOV equation  explicitly read as,
\begin{equation}\label{eq43}
\begin{split}
-\frac{d{p}}{dr}-\alpha\left[ \frac{{\nu^\prime}}{2}\,(\theta^t_t-\theta^r_r)-\frac{d \theta^r_r}{dr}+\frac{2}{r}\,(\theta^\varphi_\varphi-\theta^r_r)\right]-\frac{{\nu^\prime}}{2}\,({\rho}+{p})&\\+{\frac {\chi}{2(4\pi+\chi)}}\frac{d}{dr}\left(\rho-p\right)=0.
\end{split}
\end{equation}

the above equation says that the system is under the action of the gravitational, hydrostatic, anisotropic and $f(R,T)$ gradients. Due to the system possess spherical symmetry these gradients act along the radial direction. Thus, the compact configuration will be in hydrostatic equilibrium only if the modified TOV equation is satisfied. In Fig. \ref{f12}, we have demonstrated the evolution of the aforementioned gradients inside the object with respect to the radial coordinate $r$ due to different chosen values of $\alpha$ and $\chi$ parameters. We find that the equilibrium of the compact system is achieved for all values of $\alpha$ and $\chi$ parameters in the four different scenarios viz., GR, GR+MGD, $f\left(R,T\right)$ and $f\left(R,T\right)$+MGD validating the dynamical equilibrium of the compact structure. It should be noted that the TOV equation in the framework of GR driven the hydrostatic equilibrium for spherically symmetric compact object supported by isotropic matter distribution is recovered by fixing $\alpha$ and $\chi$ equal to zero.

\section{Energy conditions}\label{sec9}
One of the many remarkable predictions of the matter distribution that constitutes heavenly bodies can be made from a huge number of material fields. In spite of knowing the constituents that portray this material substance inside the compact structure, it could be extremely intricate to depict precisely the state of the stress-energy tensor. Actually, one has a few thoughts on the behavior of the matter under extraordinary states of density and pressure. 

Nonetheless, it is crucial to verify the viability of some inequalities corresponding to the stress-energy tensor. For this purpose, we investigate these inequalities so-called energy conditions for describing physically realistic matter configuration. The corresponding energy conditions viz., the Null Energy Condition (NEC), Strong Energy Condition (SEC) and Weak Energy Condition (WEC) are defined as
\begin{eqnarray}\label{eq92}
\text{WEC} &:& T_{\mu \nu}l^\mu l^\nu \ge 0~\mbox{or}~\tilde{\rho} \geq  0,~{\rho}^{\text{(tot)}}+{p}^{\text{(tot)}}_i \ge 0\\
\text{NEC} &:& T_{\mu \nu}t^\mu t^\nu \ge 0~\mbox{or}~~~~~~~ {\rho}^{\text{(tot)}}+{p}^{\text{(tot)}}_i \geq  0 \label{eq93}\\
&& \mbox{where}~~T_{\mu \nu}l^{\mu} \in \mbox{nonspace-like vector} \nonumber\\
\text{SEC} &:& T_{\mu \nu}l^\mu l^\nu - \frac{1} { 2} T^\lambda_\lambda l^\sigma l_\sigma \ge 0 ~\mbox{or}~ {\rho}^{\text{(tot)}}+\sum_i {p}^{\text{(tot)}}_i \nonumber\\
&& \hspace{5.5cm} \ge 0.\label{eq94}~~~~~~
\end{eqnarray}

where $i\equiv (radial~r, transverse ~t),~l^\mu$ and $t^\mu$ are time-like vector and null vector respectively. 

For the physical validity corresponding to a well-characterized stress-energy tensor, the compact stellar structure must be reliable with the inequalities (\ref{eq92})-(\ref{eq94}) simultaneously, in order to fulfill the energy conditions.
In Fig. \ref{EC}, we have shown that our stellar model is consistent with all the energy conditions and hence confirms that the physical acceptability of compact stellar interior solution.\\
Depending on the above energy conditions and its behavior, the energy must be well-defined and have an interesting result to a  reasonable physical and geometric explanation. From the physical perspective, WEC suggests that the energy density estimated by an eyewitness traversing a timelike bend is positive. NEC implies that an eyewitness crossing a null bend will quantify the surrounding energy density to be non-negative. SEC indicates that the trace of the tidal tensor estimated by the relating eyewitnesses is consistently positive.

Furthermore, the exploration about the availability of these constraints has been a source fundamental for numerous relativistic astrophysicists. It has been seen that steadiness phases of these energy conditions could help enough to investigate the steady field of heavenly structures. It is outstanding that relativistic configurations are combined with matter distributions which is depicted by its energy-momentum tensor. So as to define the arbitrariness of these tensors, it is advisable to intervene in a realistic form of matter domain. Those energy-momentum tensors that obey energy conditions, could be viewed as realistic ones. Moreover, the satisfaction of these energy conditions imposes strong limitations on the most extreme conceivable bound of the surface redshift $Z_{s}$ of the compact configuration model when the stellar system inside presents anisotropies. These suggestions will be examined in more subtleties in the following section.

\begin{figure}[!t]
\centering
\includegraphics[width=8cm,height=6cm]{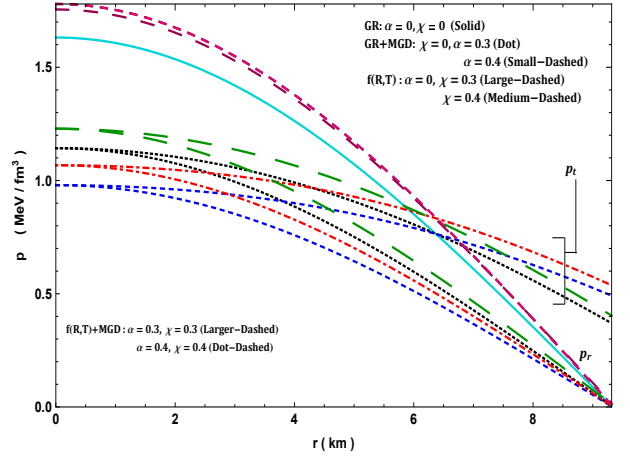}
\caption{Variation of pressures with radial coordinate $r$ for the star 4U 1820-30. For plotting of this graph, the numerical values for the constant parameters are $A = 0.001,~ D = 0.1,~ C = 0.1315$.}\label{pres}
\end{figure}
\begin{figure}[t]
\centering
\includegraphics[width=8cm,height=6cm]{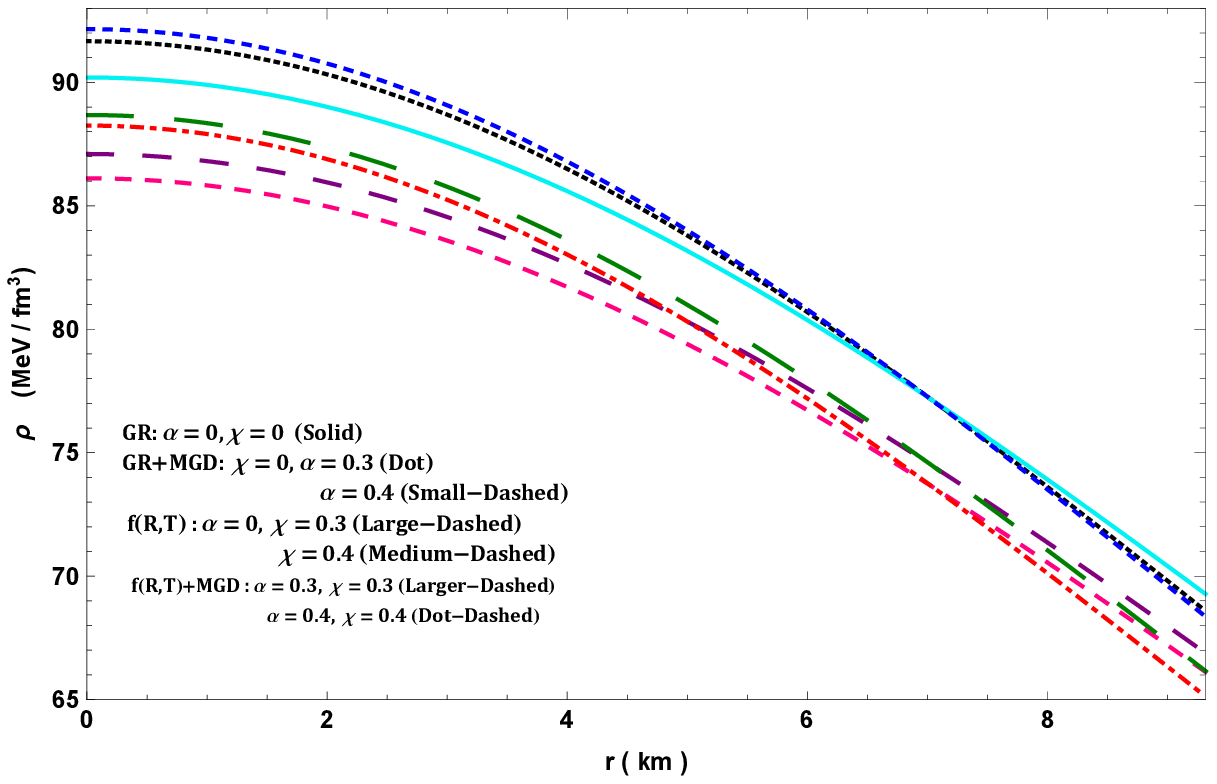}
\caption{Variation of density with radial coordinate $r$ for the star 4U 1820-30.For plotting of this graph, the numerical values for the constant parameters are $A = 0.001,~ D = 0.1,~ C = 0.1315$.}\label{dens}
\end{figure}
\begin{figure}[t]
\centering
\includegraphics[width=8cm,height=6cm]{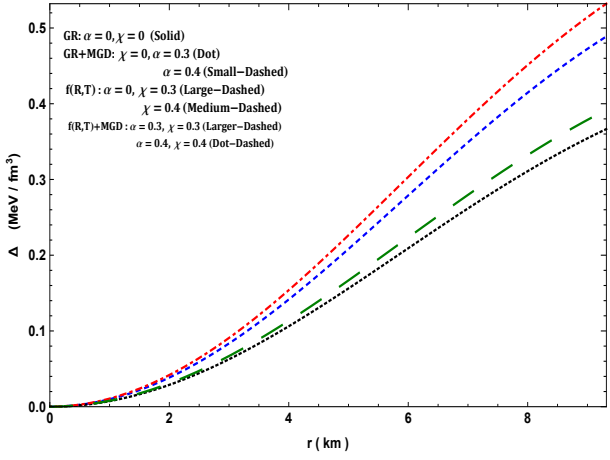}
\caption{Variation of anisotropy $\Delta$ with radial coordinate $r$ for the star 4U 1820-30.}\label{del}
\end{figure}

\begin{figure}[t]
\centering
\includegraphics[width=8cm,height=6cm]{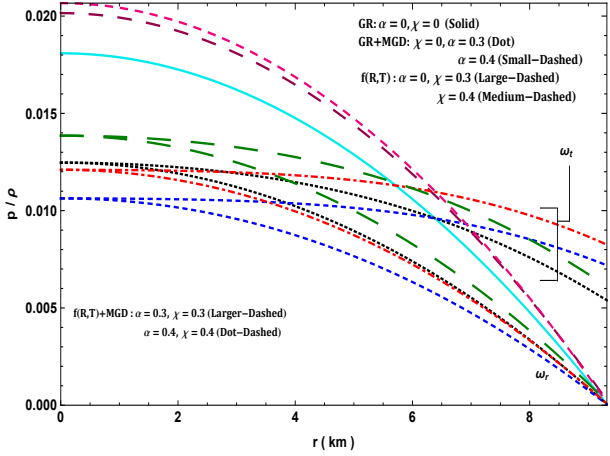}
\caption{Variation of equation of state parameters with radial coordinate $r$ for the star 4U 1820-30.}\label{eos}
\end{figure}

\section{Causality and surface redshift}\label{sec10}
We analyze the speeds of propagation related with the pressure waves in the radial and tangential directions of the compact structures as well as the modification of the upper limit of the surface redshift $Z_{s}$ as indicated by the adjustments presented by the existence of anisotropic matter distribution in the stellar system inside. The radial velocity and tangential velocity of sound (in term of total density and pressures) interior the compact structures can be achieved as,
\begin{eqnarray}
\textrm{v}^{2}_{r}=\frac{d{p}^{\text{(tot)}}_{r}}{d{\rho}^{\text{(tot)}}}~~~~ \textrm{and} ~~~~ \textrm{v}^{2}_{t}=\frac{d{p}^{\text{(tot)}}_{t}}{d{\rho}^{\text{(tot)}}}.  
\end{eqnarray}
Both the radial $(\textrm{v}_{r})$ and tangential $(\textrm{v}_{t})$ subliminal sound speed inside the compact structure must be limited by the speed of light ($c=1$ in relativistic geometrized units) in order to get a physically acceptable stellar model and hence admissibility of the resulting anisotropic solution. This leads to a so-called causality condition due to pressure waves in the fluid that do not propagate at arbitrary speeds. This condition is deterministic regardless of whether the material content of the compact structure is isotropic or anisotropic, so the only difference between them is that for the isotropic situation the subliminal sound speed, ought to be a diminishing function and in the anisotropic situation, propagation occurs in both main directions (radial and tangential directions) of the compact structure. In any case, this is not valid for the situation where there is anisotropy, since the speed behavior relies upon the inflexible nature of the material. In this vein, the causality condition reads:
\begin{equation}\label{eq43a}
0\leq \textrm{v}_{r}\leq 1 \quad \mbox{and}\quad  0\leq \textrm{v}_{t}\leq 1.  
\end{equation}
The necessary and sufficient conditions for the strong effects on the behavior of the distribution of matter within the compact stellar structure are provided in the expression (\ref{eq43a}). On the other hand, the energy-stress tensor that describes the content of the material is related to one of the conditions mentioned in inequality (\ref{eq43a}). In this respect, the stress-energy tensor is well-defined when the causality condition is conserved. Also, the reality of having various velocities affects the stability of the stellar system.

\begin{figure}[t]
\centering
\includegraphics[width=8cm,height=6cm]{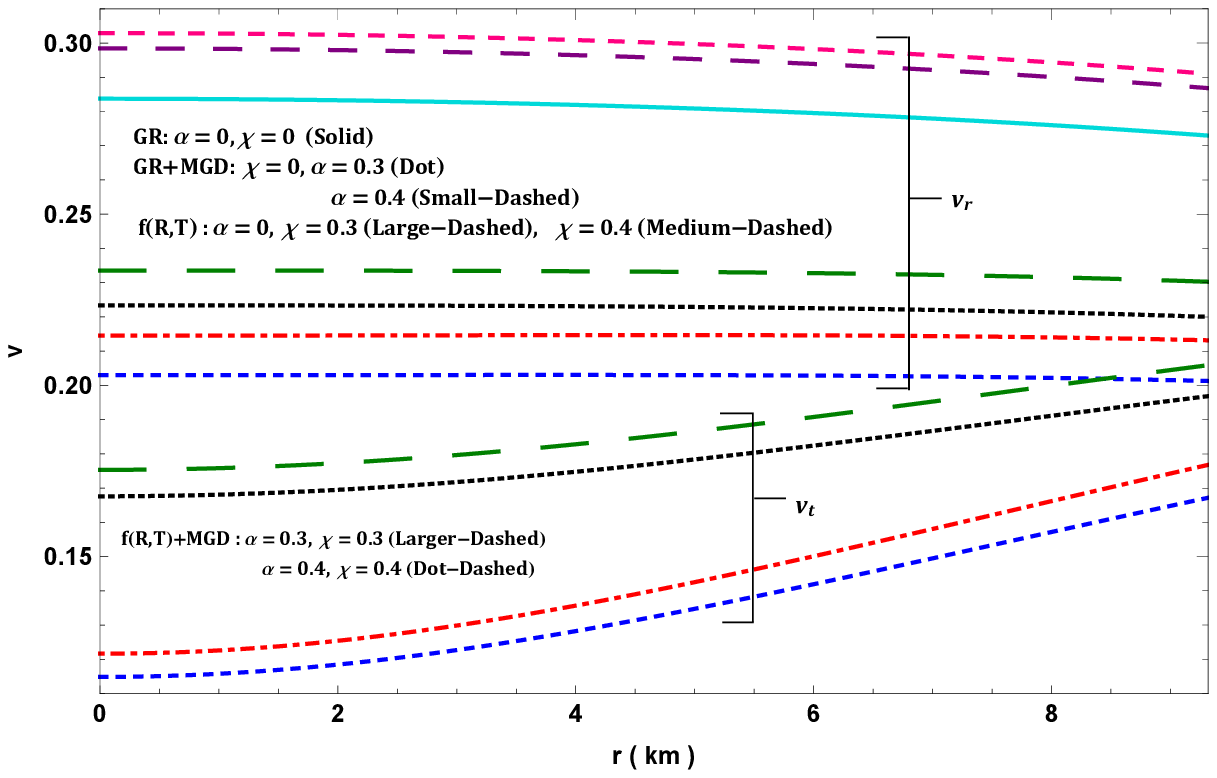}
\caption{Variation of velocity of sound with radial coordinate $r$ for the star 4U 1820-30.For plotting of this graph, the numerical values for the constant parameters are $A = 0.001,~ D = 0.1,~ C = 0.1315$.}\label{V4}
\end{figure}

In Fig. \ref{V4}, the behaviors of the radial $(\textrm{v}_{r})$ and tangential $(\textrm{v}_{t})$ sound speeds with respect to the radial coordinate $r$ for the star 4U 1820-30 has been shown and it is observed clearly that they remain inside their predetermined range ]0, 1[ throughout the stellar system, which affirm the causality condition.

The surface redshift $Z_{s}$ a noteworthy observational factor that relates the mass $\tilde{m}(R)=M$ and the radial coordinate $R$ of the compact structure, is influenced when anisotropies are brought into the stellar system, whatever the mechanism that has provoked. The factor ${2M}/{R}$ characterizes the relationship mass-radius (the compactification factor) of the stellar configuration. Buchdahl \cite{Buchdahl1959} in one of his pioneering work inferred an upper bound for the enabled mass to radius ratio, i.e., $u={2M}/{R}\leq 8/9$, for isotropic fluid spheres corresponding to a maximum value of the surface redshift equal to $2$ ($Z_{s}=2$).The relationship between $Z_{s}$, $M$ and $R$ is given in explicit form by
\begin{equation}
Z_{s}=\left(1-\frac{2M}{R}\right)^{-1/2}-1.
\end{equation}
In this respect, Bowers and Liang \cite{Bowers1974} considered an hypothetical model with incompressible fluid whose uniform density, i.e, $\rho=\rho_{0}$ and an anisotropy factor $\Delta$ with a specific form. They estimated that when the radial pressure equal to tangential pressure i.e, $p_{r}=p_{t}$ that implies the anisotropy factor is disappearing ($\Delta=0$). So, in this case, the surface redshift reached to the upper value corresponds to $Z_{s}=4.77$, and in the case, $p_{t}>p_{r}$ or $p_{t}<p_{r}$ that implies $\Delta>0$ or $\Delta<0$ the upper value of surface redshift can be exceeded. Moreover, the anisotropy factor is depending on the surface redshift which confirmed when $\Delta$ is extremely large implies $Z_{s}$ will also be too. 

\begin{figure}[!t]
\centering
\includegraphics[width=8cm,height=6cm]{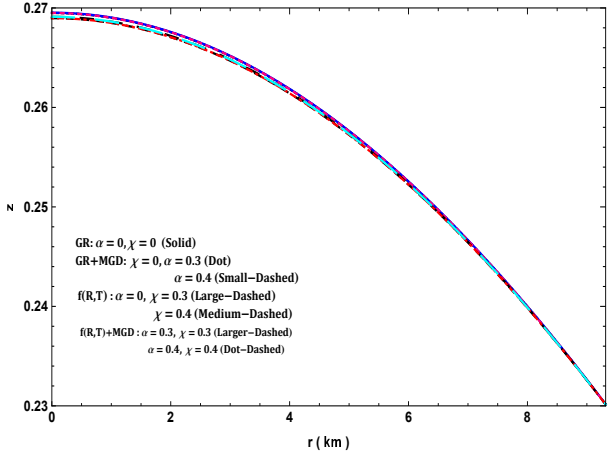}
\caption{Variation of redshift with radial coordinates.}\label{Z}
\end{figure}

In addition, Ivanov's investigations have demonstrated that for achievable anisotropic compact structure models complying with SEC and in absence of cosmological constant, the surface redshift can reach maximum higher value $Z_{s}=3.842$ though for models 
according to Ivanov \cite{Ivanov2002} in presence of cosmological constant the maximum surface redshift can go much higher up to $Z_{s}=5.211$. These qualities compare to the accompanying mass-radius relationship $0.957$ and $0.974$, separately.
The graphical representation of the surface redshift $Z_{s}$ for our stellar model is plotted in Fig. \ref{Z}. We can conclude highly obviously that it is a monotonic decreasing function with respect to the radial coordinate $r$ according to the four following scenarios, GR, GR+MGD, $f\left(R,T\right)$ and $f\left(R,T\right)$+MGD. It's value on the surface of the compact structure, i.e., $Z_{s}=0.415$ which effectively validates the approval of our stellar model as an ultra-dense compact.

\section{Adiabatic Index}\label{sec11}
The adiabatic index is a basic ingredient of the stable or unstable criteria. Therefore, the fundamental highlights of the corresponding EoS are related to the relativistic structure of the anisotropic compact configurations according to the arbitrary formulae. The stability is connected to the adiabatic index $\Gamma$, which can be explicitly composed as \cite{Chandra1964,Meraf1989,Chan1993},
\begin{equation}
\Gamma_r=\frac{\rho+p_r}{p_r}\frac{dp_r}{d\rho},
\end{equation}
where ${dp}/{d\rho}$ is the velocity of sound in units of the velocity of light. Bondi \cite{Bondi1964} provides some exceptionally valuable information for compact stellar structures and imposes some peripheral requirements. It is clearly demonstrated that a stable Newtonian perfect fluid has $\Gamma_r>4/3$ and $\Gamma =4/3$ for a neutral equilibrium. This restriction changes for a relativistic isotropic compact structure due to the correlated pressure effect, which is unsteady. Moreover, the situation turns out to be more complicated for the relativistic anisotropic compact structure, if the stability relies upon the sort of anisotropy \cite{Chan1993,Herrera1992}. In view of the above consideration, Moustakidis \cite{Moust2017} had clearly referenced that the critical value of adiabatic index strongly relies upon the mass-radius ratio ($M/R$) to tackle the issue of stability. Furthermore, the critical value is given in the explicit form as follows 
\begin{equation}
\Gamma_{crit} = \frac{4}{3} + \frac{19 }{42}~\frac{2M}{R}.
\end{equation}

\begin{figure}[!t]
\centering
\includegraphics[width=8cm,height=6cm]{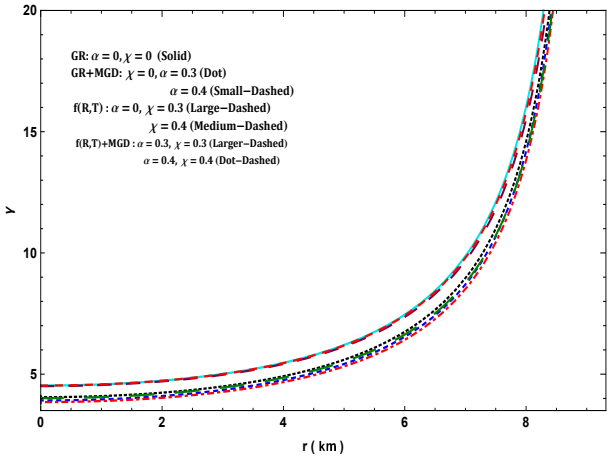}
\caption{Variation of adiabatic index with radial coordinates.}\label{Gam10}
\end{figure}

In Fig. \ref{Gam10} we have showed the behavior of the relativistic adiabatic index versus the radial coordinate $r$ of the star 4U 1820-30 for the  scenarios GR, GR+MGD, $f\left(R,T\right)$ and $f\left(R,T\right)$+MGD, which exhibits that in the case the value of the adiabatic index $\Gamma$ is greater than $4/3$ everywhere in the stellar model, confirming that our stellar system is completely stable.

\begin{figure}[t]
\centering
\includegraphics[width=8cm,height=6cm]{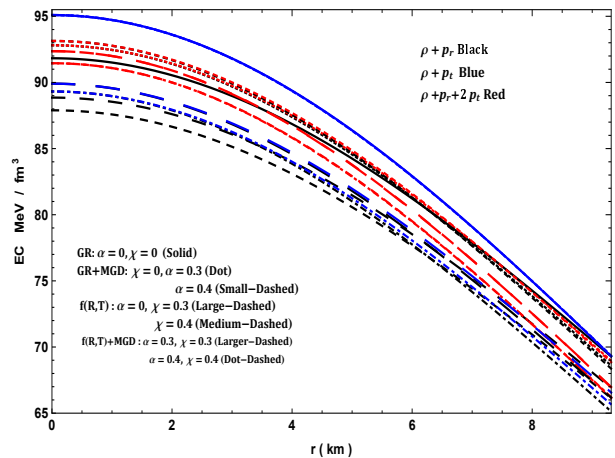}
\caption{Variation of energy conditions with radial coordinate $r$ for the star 4U 1820-30.For plotting of this graph, the numerical values for the constant parameters are $A = 0.001,~ D = 0.1,~ C = 0.1315$.}\label{EC}
\end{figure}



\section{Fitting of observed values in $M-R$ curves}\label{sec12}

We have generated the $M-R$ curves from the solution in four different scenarios i.e. GR, GR+MGD,  $f(R,T)$ and $f(R,T)+$MGD. The pure GR and $f(R,T)$ cases are isotropic while the remaining cases are anisotropic. In this model, one must keep in mind that we have assumed isotropic stress-tensor, however, due to  minimal deformation of the spacetime leads to the generation of anisotropy. Since the presence of anisotropy introduced an anisotropic force term in the modified TOV-equation, one can expect that $M-R$ curve is strongly influence by the anisotropic term. It is well known that for $\Delta > 0$, the anisotropic force is outward and  inward if $\Delta<0$. The resulting anisotropy from the solution is found to be positive thereby the anisotropic force will be radially outward. Therefore, for the case of pure GR and $f(R,T)$ the absence of anisotropic force term will render decrease in stiffness of the corresponding equation of states (EoSs) the reduces the maximum mass of the stellar fluid. Hence, one can expect that in the presence of gravitational decoupling (i.e. non-vanishing anisotropic term) the maximum mass in $M-R$ curve should also increase. For GR case i.e. $\alpha=0=\chi$, the maximum mass in $M-R$ curve is only about $1.23M_\odot$ and therefore, less massive compact stars (i.e. $M\sim 1M_\odot$) can be fitted. We have found good fitting for three compact stars namely, LMC X-4, Her X-1 and SAX J1808.4-3958. Accounting the MGD in GR the presence of anisotropic term widen the range of $M_{max}$ by changing the suitable values of decoupling paramter $\alpha$, therefore, one can fit many compact stars. In this work we have fitted for six compact stars, LMC X-4, Her X-1, SAX J1808.4-3958, PSR J1614-2230, Cyg X-2 and 4U 1538-52. The large values of $\alpha$ are suitable for fitting massive compact stars such as PSR J1614-2230.
For the case of pure $f(R,T)-$gravity, again due to the absence of anisotropic force makes the EoSs soft thereby the maximum mass is small ($\sim 1.18M_\odot$). Because of this reason, we are able to fitting only three less massive compact stars SAX J1808.4-3958, Her X-1 and 4U 1538-52. For this case, as $\chi$ increases the maximum mass decreases. Of course, in $f(R,T)+$MDG we can set for larger values of maximum mass by choosing suitable values of $\alpha$ and $\chi$. In the present work, we have fitting six massive neutron stars namely, PSR J1614-2230, Vela X-1, PSR J1903+327, 4U 1820-30, Cen X-3 and 4U 608-52 and many more can be fitted.
\begin{figure}[!t]
\centering
\includegraphics[width=8.3cm,height=6cm]{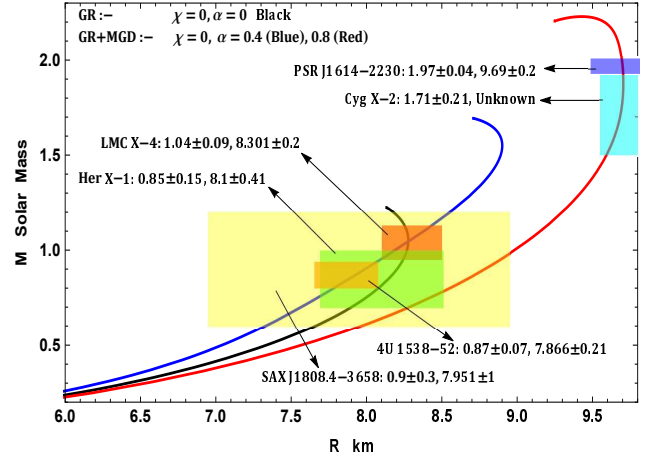}
\caption{$M-R$ curve for GR and GR+MGD using $A = 0.001,~ D = 0.1,~ C = 0.1315$ and $\rho_s=2.02 \times 10^{13}~g/cm^3$.}\label{MR7}
\end{figure}
\begin{figure}[!t]
\centering
\includegraphics[width=8.3cm,height=6cm]{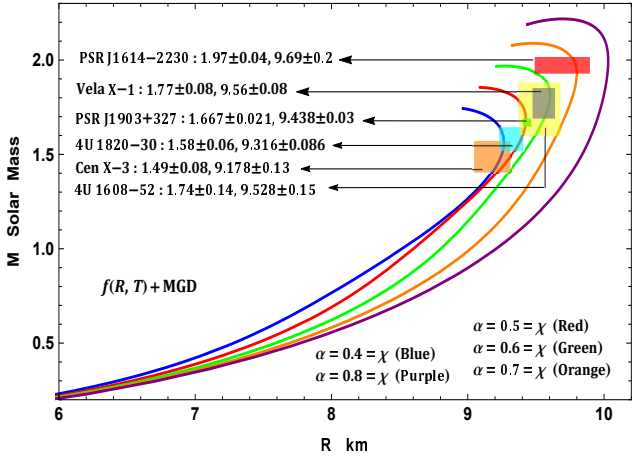}
\caption{$M-R$ curve for  $f(R,T)+MGD$ using $A = 0.001,~ D = 0.5,~ C = 0.1373$ and $\rho_s=1.6164 \times 10^{13}~g/cm^3$.}\label{MR8}
\end{figure}
\begin{figure}[!t]
\centering
\includegraphics[width=8.3cm,height=6cm]{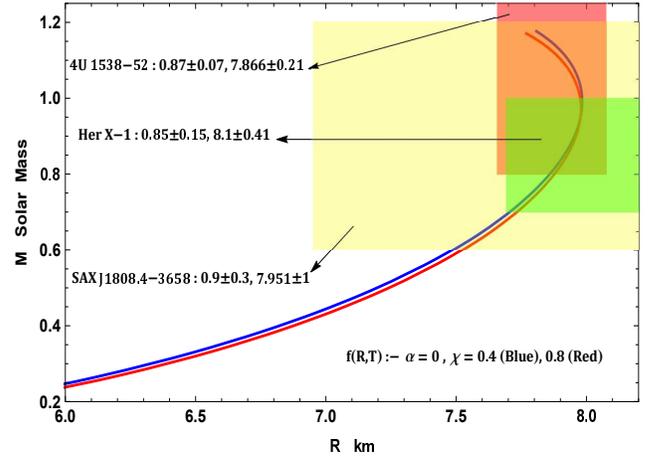}
\caption{$M-R$ curve for $f(R,T)$ using $A= 0.001,~ D = 0.5,~ C = 0.1538$ and $\rho_s=2.6941 \times 10^{13}~g/cm^3$}\label{MR9}
\end{figure}
\begin{table*}
\caption{\label{table1}Physical values of compact stellar models for the case $f(R,T)+$MGD  ($\chi=0.4=\alpha$).}
\resizebox{0.85 \hsize}{!}{$\begin{tabular}{ccccccccc}
\hline
  {Compact star}~~~~~~&{Mass}~~~~~~& Radius & Surface redshift &
Mass-radius ratio ~~~~~~& $A$~~~~~~ & $C$~~~~~~ & $D$\\
Models&$M/M_{\odot}$& $R (Km)$&($Z_s$)&$\frac{M}{R}$ &  \\ \hline
PSR J1614-2230 (Demorest\emph{et al.} \cite{Demorest2010})~~~~~~ &  1.97$\pm$ 0.04 & 9.69$\pm$ 0.2 & 0.582 & 0.300 & 0.001 & 0.1527 & 3.891   \\
Vela X-1(Rawls \emph{et al.} \cite{Rawls2011})~~~~~~ &1.77$\pm$0.08 & 9.56$\pm$ 0.08 & 0.485 & 0.273    & 0.001 & 0.1527 & 4.114  \\
PSR J1903+327 (Freire \textit{et al.} \cite{Freire2011})~~~~~~& 1.667$\pm$0.021 & 9.438$\pm$0.03 & 0.446 & 0.261 & 0.001 & 0.1527 & 4.222 \\
4U 1820-30 ~(\textrm{$G\ddot{u}ver$} \emph{et al.})  \cite{Guver2010}~~~~~~ & 1.58$\pm$0.06 & 9.316$\pm$0.086 & 0.415 & 0.250 & 0.001 & 0.1527 & 4.313  \\
Cen X-3 ~(Rawls \emph{et al.} \cite{Rawls2011}) ~~~~~~& 1.49$\pm$0.08 &9.178$\pm$0.13  & 0.386 & 0.240 & 0.001 & 0.1527 & 4.408 \\
4U 1608-52(\textrm{$G\ddot{u}ver$} \emph{et al.} \cite{Guver2010})~~~~~~& 1.74$\pm$0.14 & 9.528$\pm$0.15 & 0.473 & 0.270  & 0.001 & 0.1527 & 3.811\\ \hline
\end{tabular}$}\\
\footnotesize\textit{$^a$ In this table I, we have obtained the mass, radius, surface redshift, mass-radius ratio and constants of the different compact stars for $f(R,T)+$MGD.}
\end{table*}
\begin{table*}
\caption{\label{table2}Physical values of compact stellar models for the case GR($\alpha=0.0,\chi=0.0$) and GR+MGD ( $\alpha=0.4,\chi=0.0$).}
\resizebox{0.85 \hsize}{!}{$\begin{tabular}{ccccccccc}
\hline
  {Compact star}~~~~~~&{Mass}&~~~~~ Radius~~~~~~ & Surface redshift&~~~~
Mass-radius ratio &~~~~~~ $A$~~~~ &~~~~ $C$~~~~ &~~~~ $D$~~~~\\
Models&$M/M_{\odot}$& $R (Km)$&($Z_s$)&  $\frac{M}{R}$ &  \\ \hline
PSR J1614-2230 (Demorest\emph{et al.} \cite{Demorest2010})~~~~~~ &  1.97$\pm$ 0.04 & 9.69$\pm$ 0.2 & 0.582 & 0.30 & 0.001 & 0.1467 & 4.036 \\
Cyg X-2 (Rawls \emph{et al.} \cite{Rawls2011})~~~~~~ & 1.71$\pm$0.21 & 9.688 & 0.555 & 0.261    & 0.001 & 0.1467 & 4.370  \\
LMC X-4  (Rawls \emph{et al.} \cite{Rawls2011}) ~~~~~~& 1.04$\pm$0.09 & 8.301$\pm$0.2 & 0.260 & 0.185 & 0.001 & 0.1467 & 5.093 \\
Her X-1~(Abubekerov \emph{et al.})  \cite{Abubekerov2008}~~~~~~ & 0.85$\pm$0.15 & 8.1$\pm$0.41 & 0.204 & 0.155 & 0.001 & 0.1467 & 5.365 \\
4U 1538-52 ~(Rawls \emph{et al.} \cite{Rawls2011})~~~~~~ & 0.87$\pm$0.08 &7.866$\pm$0.21  & 0.219 & 0.163 & 0.001 & 0.1467 & 5.307    \\
SAX J1808.4-3658(Elebert \emph{et al.} \cite{Elebert2009})~~~~~~& 0.9$\pm$0.3 & 7.951$\pm$1.0 & 0.226 & 0.167 & 0.001 & 0.1467 & 4.960 \\ \hline
\end{tabular}$}\\
\footnotesize\textit{{$^b$ In this table II, we have obtained the mass, radius, surface redshift, mass-radius ratio and constants of the different compact stars for GR and GR+MGD.}}
\end{table*}

\begin{table*}
\caption{\label{table3}Physical values of compact stellar models for the case $f(R,T)$ taking $\chi=0.3$.}
\resizebox{0.85 \hsize}{!}{$\begin{tabular}{ccccccccc}
\hline
  {Compact star}~~~~~&{Mass}~~~~~& Radius~~~~~ & Surface redshift&~~~~~
Mass-radius ratio ~~~~~& $A$~~~~~ & $C$~~~~~ & $D$\\
Models&$M/M_{\odot}$& $R (Km)$&($Z_s$)&$\frac{M}{R}$ &  \\ \hline
Her X-1~(Abubekerov \emph{et al.})  \cite{Abubekerov2008}~~~~~ & 0.85$\pm$0.15 & 8.1$\pm$0.41 & 0.204 & 0.155 & 0.001 & 0.1512 & 5.215 \\
4U 1538-52 ~(Rawls \emph{et al.} \cite{Rawls2011}) ~~~~~& 0.87$\pm$0.08 &7.866$\pm$0.21  & 0.219 & 0.163 & 0.001 & 0.1512 & 5.158    \\
SAX J1808.4-3658(Elebert \emph{et al.} \cite{Elebert2009})~~~~~& 0.9$\pm$0.3 & 7.951$\pm$1.0 & 0.226 & 0.167 & 0.001 & 0.1512 & 4.812\\ \hline
\end{tabular}$}\\
\footnotesize\textit{$^c$ In this table III, we have obtained the mass, radius, surface redshift, mass-radius ratio and constants of the different compact stars for $f(R,T)$.}
\end{table*}

\section{Concluding remarks}

In this paper, we have studied the behavior of modified $f(R,T)$ gravity theory in connection to generate new exact solutions for anisotropic compact stellar configurations using the gravitational decoupling via minimal geometric deformation method. It is worth to take note that the minimal geometric deformations stem in an exclusively gravitational interaction between parts; for instance, there is no exchange of stress-energy among them. This approach has been a strategy to decouple the field equations of static and spherically symmetric self-gravitating systems. It partners the anisotropic sector with deformation over the geometrical possibilities. In this specific circumstance, the $f(R,T)$ gravity gives an enthralling point of view without including any mysterious energy component. The coupling impacts of geometry and matter components in this theory provide the non-zero covariant derivative of the trace of a stress-energy tensor which is a dominant property to talk about the traits of gravity at the quantum effect and analyzes the influence of the non-geodesic movement of test particles. The acquired solutions are explored analytically as well as graphically by acting different tests to observe the physical believability of acquired solutions.
Since $f(R,T)$ theory is not preserved,  as in other modified gravity theories \cite{Yu:2018a,Zhao:2012a} i.e., $\nabla^{\mu}T_{\mu\nu}$$\neq$$0$ new insights from the physical perspective are coming in the investigation of astrophysical interiors. In this regard, the stress-energy tensor gets a few contributions provided in the context of $f(R,T)$ theory of gravity. These contributions are managed by the additional term $\chi$ contains in the $f(R,T)$ gravity. By setting $\chi$=$0$ one recovers the usual Einstein gravity theory and therefore the conservation law of the energy-momentum tensor $\nabla^{\mu}T_{\mu\nu}$$=$$0$. For this purpose, we have investigated the impacts of this additional term and the likelihood to getting compact structures that could serve to depict quark or neutron stars. Due to the existence of this extra term, the minimal coupling matter is broken and the Bianchi identity is violated. This problem could in principle modify the mechanism of the junction conditions, as occurred in $f(R)$ gravity, for instance. In this regard, we have discussed broadly how $f(R,T)$ theory contribution stays inside the compact structure, permitting the usage of the broadest coordinating conditions namely the Darmois-Israel junction conditions. Moreover, as said earlier in Eq. (\ref{eq57}) it is an important result since the compact object will, therefore, be in equilibrium in a true exterior space-time without material content only if the total radial pressure at the surface vanishes i.e.,  the outer Schwarzschild space-time (vacuum solution). We have developed a set of singularity-free solutions of the new generalized class representing different characteristics of the anisotropic relativistic system. Moreover, we have considered the physical behavior of eleven compact spherical objects: seven X-ray binaries, namely Vela X-1, Cen X-3, Cyg X-2, LMC X-4, 4U 1538-52 and Her X-1, examined by Rawls et al. \cite{Rawls2011}; Abubekerov et al. \cite{Abubekerov2008}; Elebert et al. \cite{Elebert2009}, two low-mass-X-ray binaries, namely, 4U 1820-30 and 4U 1608-52 examined by G$\ddot{u}$ver et al. \cite{Guver2010} and two binary millisecond pulsars, namely PSR J1614–2230 and PSR J1903+327 examined by Demorest et al. \cite{Demorest2010}; Freire et al. \cite{Freire2011}, by showing the anisotropic effects presented by the $\theta$-sector effects in the $f(R,T)$ gravity theory framework. Some of the discussed properties of our systems are depicted below. 

\begin{itemize}

   \item In this study, we see that the behavior of energy density, radial pressure and tangential pressure namely,  $\rho$, $p_{r}$ and $p_{t}$ with respect to the radius $r$ for the spherical object 4U 1820-30, are appeared in  Figs. \ref{pres} and \ref{dens}, which demonstrate that all the physical amounts have their maximum values at the center diminishing continuously to arrive at the minimum values at the boundary of the spherical object. It shows that the core of the spherical object is highly compact and our stellar model is totally free of physical and geometrical singularities, which is valid for the all-area inside the spherical object.
    
    \item Moreover, the anisotropy of our stellar system is addressed in Fig. \ref{del}, which displays that the anisotropy $\Delta$ is monotonically increasing with the radial coordinate $r$ of the spherical object 4U 1820-30. For example, the anisotropy is minimum and take the zero value at the core and maximum on the surface for different values of $\alpha$ and $\chi$ of our stellar system as well as such behavior guarantee $p_{t}>p_{r}$ i.e., $\Delta>0$ wherever inside the spherical object and furthermore upgrades the stability and equilibrium mechanisms. We remark from Fig. \ref{eos} that the behavior of the equation of state parameters i.e., ($\omega_{r}=p_{r}/\rho$) and ($\omega_{t}=p_{t}/\rho$) with respect to the radial coordinate $r$,  has resulted in the content of the material inside the compact structures as expected in some senses in this type of matter distribution. This component ensures the conservation of the causality condition that leads to the physical validity of the matter distribution.

 \item On the other hand in Fig. \ref{f12}, we have exhibited the behavior of the four gradients namely, the hydrodynamic ($F_{h}$), gravitational($F_{g}$), anisotropic ($F_{a}$)  gradients, and extra gradient ($F_{frt}$) arises due to $f(R,T)$ gravity with respect to the radial coordinate $r$ due to various chosen values of $\alpha$ and $\chi$ parameters. We find that the equilibrium of these forces is achieved, which confirms that the stellar system is in dynamical equilibrium. Furthermore, in Tables \ref{table1}, \ref{table2} and \ref{table3}, we have obtained the radius, surface redshift, mass-radius ratio and the possible values of the physical parameters $A$, $C$ and $D$ of the different compact stars for various chosen values of $\alpha$ and $\chi$ parameters under four different scenarios, viz., GR, GR+MGD, $f(R,T)$ and $f(R,T)+$MGD. We have note also that the surface redshift is increasing with mass-radius ratio. 

\item We have checked that the well-behaved and positive defined stress-energy tensor wherever inside the compact spherical object which fulfill simultaneously by the inequalities (\ref{eq92})-(\ref{eq94}). In Fig. \ref{EC}, we have depicted the L.H.S of the inequalities (\ref{eq92})-(\ref{eq94}) which verifies that all the energy conditions are fulfilled at the stellar interior and hence affirms that the physical availability of the compact stellar interior solution.
\item In Fig. \ref{V4}, the practices of the radial ($v_{r}$) and tangential ($v_{t}$) sound speeds with respect to the radial coordinate $r$ for the compact structure has appeared and it is observed obviously that they stay inside their foreordained range ]0, 1[ all through the stellar system, which assert the causality condition and also confirm the admissibility of the resulting anisotropic solution of our stellar model. \item We have shown the gravitational redshift with the compact spherical object in Fig. \ref{Z}. From this figure, it is found that the surface redshift within standard estimation findings by Ivanov's \cite{Ivanov2002} which effectively validates the approval of our stellar model as an ultra-dense compact. In Fig. \ref{Gam10}, we have exhibited the behavior of the relativistic adiabatic index with respect to the radial coordinate $r$. Then we found that the value of the adiabatic index $\Gamma$ is greater than $4/3$ everywhere in the stellar structure which is affirming that our stellar system is totally stable under radial perturbations in all interior points of the astrophysical object. 
    
    \item Further, we have generated the $M-R$ curves from the solution in four different scenarios, specifically GR, GR+MGD,  $f(R,T)$ and $f(R,T)+$MGD,  which is outlined in Figs. \ref{MR7}, \ref{MR8} and \ref{MR9}. So from these three Figs. \ref{MR7}, \ref{MR8} and \ref{MR9} we have found good fitting for many compact stars in different scenarios as follows : six compact stars viz., PSR J1614-2230, Cyg X-2, LMC X-4, Her X-1, 4U 1538-52 and SAX J1808.4-3658 in both scenarios GR and GR+MGD (see Fig. \ref{MR7}), six compact stars viz., PSR J1614-2230, Vela X-1, PSR J1903+327, 4U 1820-30, Cen X-3 and 4U 1608-52 in scenario $f(R,T)+$MGD (see Fig. \ref{MR8}), and three compact stars in scenario $f(R,T)$ namely, 4U 1538-52, SAX J1808.4-3658 and Her X-1 (see Fig. \ref{MR9}) by changing the suitable values of $\alpha$ and $\chi$. In this respect, all solutions present a positive anisotropy factor  $\Delta$. In fact, this feature implies that our stellar model is completely undergoing stable behavior.
\end{itemize}
We have described the evident advantages of gravitational decoupling via minimal geometric deformation method as an efficient approach into a modified $f(R,T)$ gravity theory. This was accomplished for two distinct conditions on the gravitational source. We found new solutions are well-behaved from the physical and mathematical point of view as well as free of any geometrical singularities in the case where the mimic constraint for density and pressure. In order to compare the results obtained in the four different scenarios presented viz., GR, GR+MGD, $f(R,T)$ and $f(R,T)+$MGD, we have specified the seed solution which must satisfy the eq. (\ref{eq34}). Therefore, we take Korika-Orlyanskii perfect fluid solution as a seed solution, we came to the following conclusion: the salient radial and tangential pressure in the $f(R,T)+$MGD scenario are dominated by GR, RG+MGD and $f(R,T)$, only the salient the total energy density in the $f(R,T)+$MGD governs all situations, which suggests that the $f(R,T)+$MGD scenario is denser than GR, GR+MGD and $f(R,T)$. Apart from this, the expansion of energy density doesn't reflect an adjustment in the complete spherical object mass. Despite the fact that these solutions have the most extreme central energy density,  the $M-R$ relationships and the surface gravitational redshift undergo changes which, as examined, have strong observational ramifications.

As a final comment, we need to feature two things. To begin with, it is conceivable to acquire respectful stellar insides in the framework of the gravitational $f(R,T)$ system by utilizing gravitational decoupling via minimal geometric deformation technique. All solutions found in this paper fulfill and share all the physical and mathematical characteristics required in the investigation of compact spherical objects, which serve to comprehend the evolution of real stellar objects such as quark or neutron stars. Second, it was discovered that $f(R,T)$ gravity theory is a hopeful scenario to contemplate the existence of compact configurations depicted by an anisotropic matter distribution, which results can be compared with the well-presented general relativity, and meets the well-known and tested general requirements.

\begin{acknowledgments}
S. K. Maurya acknowledge continuous support
and encouragement from the administration of University of Nizwa. F. Tello-Ortiz thanks the financial support by the CONICYT PFCHA/DOCTORADO-NACIONAL/2019-21190856 and projects ANT-1856 and SEM 18-02 at the Universidad de Antofagasta, Chile. F. Tello-Ortiz is partially supported by grant Fondecyt No. $1161192$, Chile.
\end{acknowledgments}


\begin{thebibliography}{99}
\bibitem{Einstein:1915a}
  A.~Einstein, 
  Sitzungsber. Preuss. Akad. Wiss, {\bf 44}, 778 (1915); ibid. {\bf  48}, 844 (1915) 

  \bibitem{Wheeler:1962a}
  J.~A.~Wheeler,
  Academic, New York, pp. 25 (1962)

\bibitem{Spergel:2007a}
  D.~N.~Spergel {\it et al.} [WMAP Collaboration], 
  Astrophys. J. Suppl. {\bf 170}, 377 (2007)

\bibitem{Caldwell:2002a}
  R.~R.~Caldwell,
  Phys. Lett. B  {\bf 545}, 23 (2002)

\bibitem{Nojiri:2003a}
  S.~Nojiri and S.D.~Odinstov,
  Phys. Lett. B  {\bf 562}, 147 (2003)

\bibitem{Riess:2004nr}
  A.~G.~Riess {\it et al.} [Supernova Search Team],
  Astrophys. J.  {\bf 607}, 665 (2004)

  \bibitem{Eisenstein:2005a}
  D.J.~Eisenstein {\it et al.},
  Astrophys. J.  {\bf 633}, 560 (2005)

\bibitem{Nojiri:2003ab}
  S.~Nojiri and S.D.~Odinstov,
  Phys. Lett. B  {\bf 565}, 1 (2003)

  \bibitem{Errehymy:2017}
   A.~Errehymy {\it et al.},
   Eur. Phys. J. Plus {\bf  132}, 497 (2017)

\bibitem{Errehymy:2019}
   A.~Errehymy and M.~Daoud,
   Mod. Phys. Lett. A  {\bf  34}, 1950325 (2019)

\bibitem{Astier:2006a}
   P.~Astier {\it et al.},
   Astron. Astrophys. {\bf  447}, 31 (2006)
   
   \bibitem{Kamenshchik:2001a}
  A.~Kamenshchik {\it et al.},
  Phys. Lett. B  {\bf 511}, 265 (2001).

\bibitem{Padmanabhan:2002a}
  T.~Padmanabhan and T.~R.~Chaudhury,
  Phys. Rev. D  {\bf 66}, 081301 (2002).

\bibitem{Bento:2002a}
  M.~C.~Bento  {\it et al.},
  Phys. Rev. D {\bf 66}, 043507 (2002).

\bibitem{Utiyama:1962a}
  R. Utiyama and B. S. Dewitt, 
  J. Math. Phys. {\bf 3}, 608 (1962)

  \bibitem{Vilkovisky:1992a}
  G. A. Vilkovisky,
  Class. Quant. Gravit. {\bf 9}, 895 (1992)

\bibitem{Birrell:1982ab}
  N. D. Birrell and P. C. W. Davies, 
  Cambridge University Press, Cambridge, (1982)

\bibitem{Buchbinder:1992a}
  I. L. Buchbinder {\it et al.} 
  IOP Publishing, Bristol, (1992)

\bibitem{Pavlovic:2017ab}
  P. Pavlovic and M. Sossich,
  Phys. Rev. D  {\bf 95}, 103519 (2017)

\bibitem{Copeland:2006ab}
  E. J. Copeland {\it et al.},
  Int. J. Mod. Phys. D  {\bf 15}, 1753 (2006)

  \bibitem{Capozziello:2008a}
  S. Capozziello {\it et al.},
  Class. Quantum Gravity  {\bf 25}, 085004 (2008)

\bibitem{Capozziello:2009a}
  S. Capozziello {\it et al.},
  Mon. Not. Roy. Astron. Soc. {\bf 394}, 947-959 (2009)

\bibitem{Nojiri:2009a}
  S. Nojiri {\it et al.},
  Phys. Lett. B {\bf 681}, 74-80 (2009)

  \bibitem{Felice:2010a}
  A. de Felice and S. Tsujikawa,
  Living Rev. Relativ. {\bf 13}, 3 (2010)

\bibitem{Maartens:2010a}
  R. Maartens and R. Durrer, 
  Cambridge University Press, Cambridge, (2010)
  
\bibitem{Capozziello:2010a}
  S. Capozziello {\it et al.} 
  Class. Quant. Grav. {\bf 27}, 165008 (2010)

\bibitem{Capozziello:2011a}
  S. Capozziello
  Phys. Rev. D  {\bf 83}, 064004 (2011)

\bibitem{Nojiri:2011a}
  S. Nojiri and S. D. Odintsov,
  Phys. Rep.  {\bf 505}, 59-144 (2011)

  \bibitem{Capozziello:2012a}
  S. Capozziello {\it et al.},
  Gen. Relativ. Gravit.  {\bf 44}, 1881-1891 (2012)

  \bibitem{Astashenok:2013a}
  A. V. Astashenok {\it et al.},
  J. Cosmol. Astropart. Phys.  {\bf 2013}, 040 (2013)
  
    \bibitem{Astashenok:2015a}
  A. V. Astashenok {\it et al.},
  Phys. Lett. B  {\bf 742}, 160-166 (2015)
    
    \bibitem{Capozziello:2015a}
  S. Capozziello {\it et al.},
  Scholarpedia  {\bf 10}, 31422 (2015)
  
  \bibitem{Astashenok:2015b}
  A. V. Astashenok {\it et al.} ,
  Astrophys. Space Sci. {\bf 355}, 333-341 (2015)
  
  \bibitem{Jovanovic:2016a}
  V. B. Jovanovic {\it et al.},
  Physics of the Dark Universe {\bf 14}, 73-83 (2016)
  
  \bibitem{Capozziello:2016b}
  S. Capozziello {\it et al.},
  Phys.Rev.D {\bf 93}, 023501 (2016)
  
   \bibitem{Santos:2017a}
  C. S. Santos {\it et al.},
  Gen. Relativ. Gravit. {\bf 49}, 50 (2017)

  \bibitem{Astashenok:2017a}
  A. V. Astashenok {\it et al.},
  Class. Quant. Grav. {\bf 34}, 205008 (2017)
  
\bibitem{Chervon:2018a}
  S. V. Chervon {\it et al.},
  Nuclear Physics B {\bf 936}, 597-614 (2018)   

\bibitem{Capozziello:2018c}
  S. Capozziello {\it et al.}, 
  Phys. Lett. B {\bf 781}, 99-106 (2018)

\bibitem{Capozziello:2018b}
  S. Capozziello {\it et al.},
  arXiv preprint arXiv:1810.03204 (2018)

\bibitem{Capozziello:2018a}
  S. Capozziello {\it et al.}, 
  Phys. Lett. B {\bf 781}, 99-106 (2018)

  \bibitem{Odintsov:2019a}
  S. D. Odintsov and V. K. Oikonomou,
  Phys. Rev. D  {\bf 99}, 064049 (2019)

 \bibitem{Odintsov:2019b}
  S. D. Odintsov and V. K. Oikonomou,
  Class. Quant. Grav.  {\bf 36}, 065008 (2019)

\bibitem{Capozziello:2019a}
  S. Capozziello and R. D’Agostino, 
  Gen. Relat. Gravit. {\bf 51}, 2 (2019)
  
  \bibitem{Bohmer:2011a}
  C. G. B$\ddot{o}$hmer {\it et al.},
  Class. Quant. Grav.  {\bf 28}, 245020 (2011)
    
    \bibitem{Wang:2011a}
  T. Wang,
  Phys. Rev. D {\bf 84}, 024042 (2011)
  
  \bibitem{Daouda:2011a}
  M. H. Daouda {\it et al.},
  Eur. Phys. J. C  {\bf 71}, 1817 (2011)
  
    \bibitem{Sharif:2013a}
  M. Sharif and S. Rani,
  Phys. Rev. D {\bf 88}, 123501 (2013)  

\bibitem{Capozziello:2013b}
  S. Capozziello {\it et al.}, 
  JHEP {\bf 2013}, 039 (2013)   
    
    \bibitem{Harko:2011a}
  T. Harko {\it et al.},
  Phys. Rev. D {\bf 84}, 024020 (2011)

    \bibitem{pop}
  N. J. Poplawski, arXiv:gr-qc/0608031


    \bibitem{Yousaf:2016a}
  Z. Yousaf {\it et al.},
  Phys. Rev. D  {\bf 93}, 124048 (2016)
  
  \bibitem{Yousaf:2016b}
  Z. Yousaf {\it et al.},
  Phys. Rev. D  {\bf 93}, 064059 (2016)  
    
    \bibitem{Correa:2016a}
  R. A. C. Correa and P. H. R. S. Moreas,
  Eur. Phys. J. Plus  {\bf 76}, 100-106 (2016)
  
    \bibitem{Moraes:2017a}
  P. H. R. S. Moraes {\it et al.},
  arXiv: 1701.01027v1.

    \bibitem{Das:2016a}
  D. Das {\it et al.},
  Eur. Phys. J. C  {\bf 76}, 654 (2016)

    \bibitem{Deb:2018a}
  A. Deb {\it et al.},
  JCAP  {\bf 1503}, 044 (2018)

  \bibitem{Maurya:2019ab}
  S. K. Maurya {\it et al.},
  Phys. Rev. D  {\bf 100}, 044014 (2019)  

    \bibitem{Deb:2019a}
  A. Deb {\it et al.},
  Mon. Not. Roy. Astron. Soc.  {\bf 485}, 5652 (2019)

    \bibitem{Sahoo:2019a}
  P. K. Sahoo {\it et al.},
  Int. J. Mod. Phys. D  {\bf 28}, 1950004 (2019)

    \bibitem{Shabani:2018a}
  H. Shabani and A. Hadi Ziaie,
  Eur. Phys. J. C  {\bf 78}, 397-445 (2018)
  
    \bibitem{Wu:2018a}
  J. Wu {\it et al.},
  Eur. Phys. J. C  {\bf 78}, 1-22 (2018)

    \bibitem{Barrientos:2018a}
  E. Barrientos {\it et al.},
  Phys. Rev. D  {\bf 97}, 104041 (2018)

    \bibitem{Hansraj:2018a}
  S. Hansraj and A. Banerjee,
  Phys. Rev. D  {\bf 97}, 104020 (2018)

    \bibitem{Singh:2018a}
  J. K. Singh {\it et al.},
  Phys. Rev. D  {\bf 97}, 123536 (2018)

    \bibitem{Baffou:2018a}
  E. H. Baffou {\it et al.},
  arXiv: 1808.01917 (2018)

    \bibitem{Hansraj:2018b}
  S. Hansraj,
  Eur. Phys. J. C  {\bf 78}, 700-707 (2018)

    \bibitem{Yousaf:2018b}
  Z. Yousaf {\it et al.},
  Eur. Phys. J. C  {\bf 78}, 307-333 (2018)

    \bibitem{Bengochea:2009a}
  G.R. Bengochea and R. Ferraro,
  Phys. Rev. D  {\bf 79}, 124019 (2009)

    \bibitem{Linder:2010b}
  E. V. Linder,
  Phys. Rev. D  {\bf 81}, 127301 (2010)

    \bibitem{Momeni:2014a}
  D. Momeni and R. Myrzakulov,
  Int. J. Geom. Meth. Mod. Phys.  {\bf 11}, 1450077 (2014)

    \bibitem{Junior:2015a}
  E.L.B. Junior {\it et al.},
  Class. Quant. Gravit. {\bf 33}, 125006 (2015)

    \bibitem{Nassur:2015a}
  S.B. Nassur {\it et al.},
  Astrophys. Space Sci. {\bf 360}, 60 (2015)

    \bibitem{Salako:2015a}
  I.G. Salako {\it et al.},
  Astrophys. Space Sci. {\bf 358}, 13 (2015)

    \bibitem{Saez-Gomez:2016a}
  D. Saez-Gomez {\it et al.},
  Phys. Rev. D {\bf 94}, 024034 (2016)

    \bibitem{Pace:2017a}
  M. Pace and J. L. Said,
  Eur. Phys. J. C {\bf 77}, 62 (2017)
 \bibitem{Lobo1}T.~Harko, F.~S.~N.~Lobo, G.~Otalora and E.~N.~Saridakis,
  JCAP {\bf 1412}, 021 (2014)
  doi:10.1088/1475-7516/2014/12/021
  [arXiv:1405.0519 [gr-qc]]. 
 \bibitem{Lobo2} T.~Harko, F.~S.~N.~Lobo, G.~Otalora and E.~N.~Saridakis,
  Phys.\ Rev.\ D {\bf 89}, 124036 (2014)
  doi:10.1103/PhysRevD.89.124036
  [arXiv:1404.6212 [gr-qc]]. 
    \bibitem{Bamba:2010a}
  K. Bamba {\it et al.},
  Euro-phys. Lett. {\bf 89}, 50003 (2010)

    \bibitem{Bamba:2010b}
  K. Bamba {\it et al.},
 Eur. Phys. J. C {\bf 67}, 295 (2010)
  
\bibitem{Rodrigues:2014a}
  M. E. Rodrigues {\it et al.},
  Can. J. Phys.  {\bf 92}, 173 (2014)

\bibitem{Nojiri:2005a}
  S.~Nojiri and S.D.~Odinstov,
  Phys. Lett. B  {\bf 631}, 1 (2005)

\bibitem{Capozziello:2002b}
  S. Capozziello,
  Int. J. Mod. Phys. D  {\bf 11}, 483-492 (2002)

\bibitem{Carroll:2004a}
  S. M. Carroll {\it et al.},
  Phys. Rev. D  {\bf 70}, 043528 (2004)

\bibitem{Dolgov:2003a}
  A. D. Dolgov and M. Kawasaki,
  Phys. Lett. B  {\bf 573}, 1-4 (2003)

\bibitem{Nojiri:2003b}
  S.~Nojiri and S.D.~Odinstov,
  Phys. Rev. D  {\bf 68}, 123512 (2003)

\bibitem{Nojiri:2004c}
  S.~Nojiri and S.D.~Odinstov,
  Mod. Phys. Lett. A  {\bf 19}, 627-638 (2004)

\bibitem{Copeland:2006a}
  E. J. Copeland {\it et al.},
  Int. J. Mod. Phys. D  {\bf 15}, 1753-1845 (2006)

\bibitem{Starobinsky:1980a}
  A. A. Starobinsky,
  Phys. Lett. B  {\bf 91}, 99-102 (1980)

\bibitem{Starobinsky:2007a}
  A. A. Starobinsky,
  JETP Lett.  {\bf 86}, 157-164 (2007)

\bibitem{Hu:2007a}
  W. Hu and I. Sawicki,
  Phys. Rev. D  {\bf 76}, 064004 (2007)

\bibitem{Capozziello:2014c}
  S. Capozziello {\it et al.},
  Int. J. Mod. Phys. D  {\bf 24}, 1541002 (2014)

\bibitem{Cruz-Dombriz:2016a}
  A. D. l. Cruz-Dombriz {\it et al.},
  J. Cosmol. Astropart. Phys.  {\bf 1612}, 042-067 (2016)

\bibitem{Aviles:2013a}
  A. Aviles {\it et al.},
  Phys. Rev. D  {\bf 87}, 064025 (2013)

\bibitem{D'Agostino:2013a}
  R. D'Agostino and O. Luongo,
  arXiv: 1807.10167 (2018)

\bibitem{Gibbons:1977a}
  G. W. Gibbons and S. W. Hawking,
  Phys. Rev. D  {\bf 15}, 2738 (1977)

\bibitem{Rastall:1972a}
  P. Rastall,
  Phys. Rev. D  {\bf 6}, 3357 (1972)

\bibitem{Parker:1971a}
  L. Parker,
  Phys. Rev. D  {\bf 3}, 346 (1971)
  
\bibitem{Ford:1987a}
  L. H. Ford,
  Phys. Rev. D  {\bf 35}, 2955 (1987)

\bibitem{Birrell:1982a}
  N. D. Birrell and P. C. W. Davies,
  Quantum Fields in Curved Space (Cambridge University Press, Cambridge, 1982) (1982)

\bibitem{Brax:2007a}
  Ph. Brax {\it et al.},
  arXiv:0706.1024 (2007)
 
 \bibitem{Smalley:1984a}
  L. L. Smalley,
   Il Nuovo Cimento B, 80, 1, 42 (1984) (1984)

\bibitem{Oliveira:2016a}
  A. M. Oliveira {\it et al.},
  Phys. Rev. D {\bf 93}, 124020 (2016)

\bibitem{Bronnikov:2016a}
  K. A. Bronnikov {\it et al.},
  arXiv:1606.06242 (2016)

\bibitem{Moradpour:2016a}
  H. Moradpour and N. Sadeghnezhad,
  arXiv:1606.00846 (2016)
  
\bibitem{Hermano:2017} 
H. Velten and T. R. P. Carames
Phys. Rev. D {\bf95}, 123536 (2017)

\bibitem{Mak:2004a}
  M. K. Mak and T. Harko,
  Int. J. Mod. Phys. D {\bf 13}, 149 (2004)

\bibitem{Gleiser:2004a}
  M. Gleiser and K. Dev,
  Int. J. Mod. Phys. D {\bf 13}, 1389 (2004)

\bibitem{Kalam:2013a}
  M. Kalam {\it et al.},
  Eur. Phys. J. C {\bf 73}, 2409 (2013)

\bibitem{Maurya:2017a}
  S. K. Maurya {\it et al.},
  Eur. Phys. J. C {\bf 77}, 360 (2017)
  
 \bibitem{Deb:2019}
  D. Deb{\it et al.},
  M. N. R. A. S. {\bf485}, 5652 (2019)   
  
\bibitem{Maurya:2019b}
  S. K. Maurya {\it et al.},
  Phys. Rev. D {\bf 100}, 044014 (2019)  
  
  \bibitem{Karmarkar:1948}
 K.R. Karmarkar, Proc. Indian. Acad. Sci. A {\bf27}, 56 (1948) 

\bibitem{Ovalle:2008b}
  J. Ovalle,
 Mod. Phys. Lett. A {\bf 23}, 3247 (2008)
 \bibitem{Ovallebook} J. Ovalle and R. Casadio, Beyond Einstein Gravity. The Minimal Geometric Deformation Approach in the Brane-World. Cham: Springer Nature (2020). ISBN: 9783030394929 DOI:10.1007/978-3-030-39493-6

 \bibitem{Ovalleb} R. Casadio, J. Ovalle, Phys. Lett. B 715, 251 (2012).
\bibitem{Ovallec}  J. Ovalle, F. Linares, A. Pasqua, A. Sotomayor, Class. Quant. Gravit. 30, 175019 (2013). 
\bibitem{Ovalled}  R. Casadio, J. Ovalle, R. da Rocha, Class. Quant. Gravit. 30, 175019 (2014)
\bibitem{Ovallee} R. Casadio, J. Ovalle, R. da Rocha,  Euro. phys. Lett. 110, 40003 (2015).
\bibitem{Ovallef} J. Ovalle, L. A. Gergely, R. Casadio, Class. Quant. Gravit. 32, 045015 (2015).
\bibitem{Ovalleg}  R. da Rocha,  Eur. Phys. J. C 77, 355 (2017).
\bibitem{Ovallei} J. Ovalle, F. Linares, Phys. Rev. D 88, 104026 (2013).
\bibitem{Ovallej} R. Casadio, J. Ovalle, R. da Rocha, Class. Quant. Gravit. 32, 215020 (2015)
\bibitem{Ovallek} J. Ovalle, Phys. Lett. B 788, 213 (2019)

 \bibitem{Ovalle:2018c}  J. Ovalle, R. Casadio, R. da Rocha, A. Sotomayor,  Eur. Phys. J. C 78, 122 (2018) 
 \bibitem{Gabbanelli:2018}
  L. Gabbanelli, A. Rincón and C. Rubio, Eur. Phys. J. C {\bf 78}, 370 (2018)
 
 \bibitem{Tello:2018}
  M. Estrada and F. Tello-Ortiz, Eur. Phys. J. Plus. {\bf 133}, 153 (2018)
 
   \bibitem{Morales:2018aa} E. Morales and F. Tello-Ortiz, Eur. Phys. J. C {\bf 78}, 618 (2018)
  
  \bibitem{Morales:2018a}
  E. Morales and F. Tello-Ortiz,
  Eur. Phys. J. C {\bf 78}, 841 (2018)
  
\bibitem{Maurya:2019a}
  S. K. Maurya and F. Tello-Ortiz,
  Eur. Phys. J. C {\bf 79}, 85 (2019)
 
 \bibitem{Casadio:2015a}
  R. Casadio {\it et al.},
 Class. Quant. Grav. {\bf 32}, 215020 (2015)
 
 
   \bibitem{Sharif:2018c}
   M. Sharif and S. Sadiq,
 Eur. Phys. J. C {\bf 78}, 410 (2018)
 
\bibitem{fr1} P.H.R.S. Moraes, J.D.V. Arbanil and M. Malheiro, J. Cosmol. Astropart. Phys.  {\bf 2016}, 005 (2016)
\bibitem{fr2} M. Sharif and A. Siddiqa, Eur. Phys. J. Plus 132, 529 (2017)
\bibitem{fr3} G. A. Carvalho1, R. V. Lobato, P. H. R. S. Moraes, J. D. V. Arbañil et al., Eur. Phys. J. C 77, 871 (2017)
\bibitem{fr4} M. Sharif and A. Siddiqa, Eur. Phys. J. Plus 133,  226 (2018)
\bibitem{fr5} S. Biswas, S. Ghosh, S. Ray, F. Rahaman, B.K. Guha, Annals of Physics, doi.org/10.1016/j.aop.2018.12.004 (2018) 
\bibitem{fr6} G. Mustafa, M. Zubair, S. Waheed, X. Tiecheng, Eur. Phys. J. C 80, 26 (2020)

\bibitem{fr7} M. Sharif and A. Waseem, Int. J. Mod. Phys. D, DOI:10.1142/S0218271819500330 (2018)

\bibitem{fr8} A. Das, S. Ghosh, B. K. Guha,S. Das, F. Rahaman, S. Ray, Phy. Rev. D 95, 124011 (2017)

 
 
 
    \bibitem{Yu:2018a}
   H. Yu {\it et al.},
 Phys. Rev. D {\bf 97}, 083524 (2018)
 
    \bibitem{Zhao:2012a}
   Y. Y. Zhao,
 Eur. Phys. J. C {\bf 72}, 1924 (2012)
 
\bibitem{Israel:1966a}
  W. Israel,
  Nuovo Cim. B,  {\bf 44}, 1 (1966)
  
    \bibitem{Darmois:1927a}
  G. Darmois, Memorial des Sciences Mathematiques (Gauthier-Villars, Paris, 1927), Fasc. {\bf 25} (1927)
  
  \bibitem{Estrada:2019a}
  M. Estrada and R. Prado,
  Eur.Phys.J.Plus {\bf 134}, 168 (2019)
  
  
     \bibitem{Korkina:1991a}
  M. P. Korkina and O. Y. Orlyanskii,
  Ukr. J. Phys. {\bf 36}, 885 (1991), translated to english from Ukr. Fiz. Zh. {\bf 36}, 127 [K-O III VI VII] (1991)

   \bibitem{Maurya:2012a}
  S. K. Maurya, and Y. K. Gupta,
  Astrophys. Space Sci. {\bf 340}, 323 (2012)

 \bibitem{Maurya:2018aa}
S. K. Maurya {\it et al.}, 
Phys.Rev. D {\bf 97}, 044022 (2018)
  
\bibitem{Rawls2011}
M. L. Rawls, J. A. Orosz,  J. E.  McClintock,  M. A. P. Torres, C. D. Bailyn, M. M. Buxton, ApJ, 730, 25 ( 2011)

\bibitem{Freire2011}
P. C. C. Freire  et al.,  MNRAS, 412, 2763 (2011)

\bibitem{Demorest2010}
P. B. Demorest, T. Pennucci, S. M. Ransom, M. S. E. Roberts, J. W. T. Hessels, Nature, 467, 1081 (2010)
 
\bibitem{Guver2010}
T. G$\ddot{u}$ver, P. Wroblewski, L. Camarota, F. Ozel , Astrophys. J., 719, 1807 (2010)
 
\bibitem{Abubekerov2008}
M. K. Abubekerov, E. A. Antokhina, A. M. Cherepashchuk, V. V. Shimanskii, Astron. Rep., 52, 379 (2008)
 
\bibitem{Elebert2009}
P. Elebert et al., Mon. Not. R. Astron. Soc., 395, 884 (2009)

\bibitem{Buchdahl1959}
H.A. Buchdahl, Phys. Rev. 116, 1027 (1959)

\bibitem{Bowers1974}
R. L. Bowers and E. P. T. Liang, Astrophys. J. 188, 657 (1974).

\bibitem{Ivanov2002}
B.V. Ivanov, Phys. Rev. D 65, 104011 (2002)

\bibitem{Chandra1964}
S. Chandrasekhar, Astrophys. J. 140, 417 (1964)

\bibitem{Meraf1989}
M. Merafina and R. Ruffini, A $\&$ A 221, 4 (1989)

\bibitem{Chan1993}
R. Chan, L. Herrera and N. O. Santos, Mon. Not. R. Astron. Soc. 265, 533 (1993)
\bibitem{Bondi1964}
H. Bondi, Proc. R. Soc. Lond. A 281, 39 (1964)

\bibitem{Herrera1992}
L. Herrera, Phys. Lett. A, 165, 206 (1992)

\bibitem{Moust2017}
Ch. C. Moustakidis, Gen. Relativ. Gravit. 49, 68 (2017)


\end{thebibliography}
\end{document}